\documentclass[aps,pra,twocolumn]{revtex4-1}

\usepackage{amsmath}
\usepackage{amssymb}
\usepackage{graphicx}
\usepackage{braket}
\usepackage{comment}
\usepackage{hyperref}
\usepackage{soul,color}
\usepackage[bottom]{footmisc}

\renewcommand{\vec}[1]{\mathbf{#1}}
\newcommand{\erfi}{\mathrm{Erfi}}

\usepackage{float}
\begin{document}

\title{Orientational alignment in Cavity Quantum Electrodynamics}

\author{Jonathan Keeling}
\email{jmjk@st-andrews.ac.uk}
\affiliation{SUPA, School of Physics and Astronomy, University of St Andrews, St Andrews, KY16 9SS, United Kingdom}

\author{Peter G. Kirton}
\affiliation{SUPA, School of Physics and Astronomy, University of St Andrews, St Andrews, KY16 9SS, United Kingdom}
\date{\today}

\begin{abstract}
We consider the orientational alignment of dipoles due to strong matter-light
coupling, for a non-vanishing density of excitations.  We compare
various approaches to this problem in the limit of large numbers of emitters,
and show that direct Monte Carlo integration, mean-field theory, and 
large deviation methods match exactly in this limit.   All three results
show that orientational alignment develops in the presence of a macroscopically
occupied polariton mode, and that the dipoles asymptotically approach
perfect alignment in the limit of high density or low temperature.
\end{abstract}

\maketitle

\section{Introduction}

When light couples to matter strongly enough, it can change material
properties.  This general idea has recently seen an explosion of interest
across a variety of materials and for a range of physical phenomena, as
reviewed briefly below.  The most dramatic such effects occur when matter-light
coupling induces a phase transition, leading to changes of
material properties.  Phase transitions occur in the thermodynamic
limit, and so rely on understanding matter-light coupling with large
numbers of particles. It is therefore important to test approximate
theoretical methods that describe matter-light coupling in this limit.
Here we provide a comparison of two such methods, mean-field theory and
large deviation approaches, in the context of orientational ordering of
dipoles coupled to light.

One context in which changes to material properties due to matter-light
coupling have been extensively studied is that of organic molecules,
which already have interesting photophysics and chemistry even without
strong coupling~\cite{AgranovichRussian,Agranovich2009a,barford13}.
In particular, the possibility to manipulate chemical
reaction rates, or allow photocatalysis of multiple reactions by a single
photon~\cite{Schwartz11,Hutchison12,Herrera15,Galego2016,Kowalewski2016,Flick2017a,galego17}
has been studied in such materials.  Similarly, the idea of modifying
electrical transport~\cite{orgiu15,Feist2014,Hagenmuller2018} by strong
coupling to a cavity has also been explored. Another developing area is in using strong coupling
to affect singlet fission~\cite{Martinez-Martinez2017}, potentially
improving solar cell performance.  There have also been many
works exploring whether the configuration, vibrational state, or
orientation of a molecule can be affected by strong coupling to
light~\cite{Canaguier-durand2013,Shalabney15,Galego15,Cwik2016,Galego2016,Feist2018,Martinez-Martinez2018},
and how strong matter-light coupling may lead to the breakdown of the
Born-Oppenheimer approximation~\cite{Galego15,Bennett2016,Flick2017b}.
Recently, there have been several reviews discussing these developments,
see for example~\cite{Herrera2018,Kolaric2018,Ribeiro2018,Feist2018}.

In other contexts, strong driving by external light has been used as a way
to induce transient superconductivity~\cite{Fausti2011,Mitrano2016}, with a
variety of proposed
mechanisms~\cite{Mankowsky2014,Knap2016,Kennes2017,sentef17,Schlawin2017}.
Superconductivity can also be affected by strong matter-light coupling of
phonon modes to an infra-red cavity mode~\cite{Sentef2018}, or using multimode Terahertz cavities to induce cavity-mediated electron pairing~\cite{schlawin18}.  Similarly
structural phase transitions in Perovskites have been found to be modified
by strong coupling~\cite{Wang2014}.  In the context of organic molecules,
strong coupling between infra-red cavities and vibrational modes has also
been studied~\cite{George15,Pino15,shalabney15:raman,Pino15a,Shalabney15,strashko2016raman}.

For many of the above effects, a particularly interesting feature is the
possibility of collective effects --- i.e. effects of there being many
molecules coupled to the same cavity mode.  To understand such collective
effects, it is necessary to consider the behavior in the limit of a large
number, $N$, of emitters.  While for small $N$ it is possible to
consider exact numerical methods, such as adapting density functional
theory to include cavity QED~\cite{Flick2017}, such exact approaches are
challenging for macroscopic numbers of emitters.  A number of different
theoretical frameworks have been used for tackling these problems.  These
 include using symmetries to
reduce the problem size, and mean-field
theories~\cite{Herrera15,Herrera2016a,Herrera2016,Cortese2017,Zeb2017}.
From the context of condensed matter physics, mean-field theory is a natural
approach.  For  $N$ emitters coupled to a single mode, mean-field theory is expected to become exact as $N \to \infty$, with corrections scaling as $1/N$.  We have shown
elsewhere~\cite{Zeb2017} how the absorption spectra of vibrationally dressed
molecules can indeed be recovered by such an approach.  Here we consider other
forms of dressing, and the comparison between mean-field theory and
exact numerical methods.

In this article, we focus on the question, first discussed by
\citet{Cortese2017}, of how a strong coupling can lead to orientational
alignment of molecular dipoles. We compare various approaches to answering
this question, using mean-field theory~\cite{wang73,eastham01}, direct
Monte-Carlo integration, and large deviation
approaches~\cite{Touchette2009}.  We find that in the limit of large $N$,
these results all agree (when considering parameter values for
which agreement can be expected).  We also show the versatility of
mean-field approaches to include saturation effects expected at
high excitation density.  In an appendix, we also show how
these methods can be easily adapted to a wider set of related models.

\section{Model and summary of previous results}
\label{sec:model-prev-results}

We consider a model of $N$ orientable dipoles, strongly coupled to a single cavity mode.  Such a model was introduced previously in Refs.~\onlinecite{Cwik2016,Cortese2017}. The electronic states of the dipoles are modeled as two-level systems, corresponding to ground and first excited electronic states.  Such a description is appropriate when the dipole has an anharmonic spectrum, and this first electronic transition dominates the optical response.  The electronic state is thus described by Pauli matrices $\sigma_i$ and the cavity mode by the creation operator $a^\dagger$.
The coupling
strength of a dipole, depends on its orientation relative to the electric
field direction (we assume a single polarization for simplicity). For
dipoles free to rotate in two dimensions we parameterize this by a single
alignment angle, $\theta_i$.  This leads to the generalized Dicke model~\cite{Garraway2011a}:
\begin{equation}
  \label{eq:Hamil}
  H = \omega a^\dagger a
  + \sum_i g \cos(\theta_i) (a^\dagger \sigma_i^- + a^{} \sigma_i^+)
  + \frac{\omega_0}{2} \sigma_i^z.
\end{equation}
The polariton splitting emerging from such a model scales as $g\sqrt{N}$,
so in the following we assume $g\sqrt{N}$  is intensive and so remains finite in the limit of large $N$.  Physically, this scaling occurs because the matter-light coupling $g$ in Eq.~(\ref{eq:Hamil}) scales as $1/\sqrt{V}$ where $V$ is the quantization volume, so $g \sqrt{N}$ scales as the square root of the density of dipoles, an intensive quantity.

Such a model may be considered as describing the orientation of organic
molecules in solution, with strong coupling to an optical cavity mode.  We note that strong coupling between organic molecules and infra-red cavities has also been studied, however in such a case the electromagnetic mode couples to the displacement of a vibrational mode of the molecule~\cite{George15,Pino15,shalabney15:raman,Pino15a,Shalabney15,strashko2016raman}.  The model in Eq.~(\ref{eq:Hamil}), involving transitions of two-level systems,  specifically describes coupling  to electronic transitions, not vibrational modes, so we focus only on strong coupling to optical cavities. Closely related models can arise in other contexts.  For
example, there is a close connection to a model
considered in the context of cold atoms in an optical
cavity~\cite{Mivehvar2017}, where a Raman process between cavity light and
external pump can cause a change of spin state of the atoms, $\sigma_i^\pm$; 
in this case $\theta_i$ denotes the position of the atom in a standing wave of light. Similar models can also be realized in arrays of superconducting qubits~\cite{Kakuyanagi2016}.

As Eq.~(\ref{eq:Hamil}) is a modified version of the Dicke
model~\cite{Garraway2011a}, such a model can naturally be expected to
undergo a version of the Dicke-Hepp-Lieb phase transition~\cite{wang73}.
This has been extensively studied in the absence of an orientational degree of
freedom, i.e. setting $\theta_i=0$.  In particular, if one considers
Eq.~(\ref{eq:Hamil}) in the grand canonical ensemble, with a chemical
potential $\mu$ controlling the number of excitations $ M = a^\dagger a
+ \sum_i (\sigma_i^z+1)/2$, there is a transition at low temperatures or
high densities to a state where there is a macroscopic occupation of the
photon mode~\cite{eastham01}.  We will discuss further below how
this transition is modified by the orientational degree of freedom.

In
\citet{Cortese2017}, the behavior of angular orientation following
from Eq.~(\ref{eq:Hamil}) in the
$M$ excitation sector ground state. i.e., the evolution of
$\langle \cos^2 \theta \rangle$ as a function of density, $M/N$ and
temperature was studied. For reference, we summarize these results
here.  Focusing on the resonant case,
$\omega_0=\omega$, we may approximately write the energy of
the $M$ polariton states as $\epsilon_M \simeq -M g \sqrt{\sum_i \cos^2\theta_i}$, which leads to an effective partition function:
\begin{equation}
  \label{eq:exact-Z}
  \mathcal{Z} = \prod_i \int \!\!d\theta_i 
  \exp\left( \beta M g \sqrt{\sum_i \cos^2\theta_i}\right).
\end{equation}
This expression neglects any saturation of the polariton splitting at
finite excitation density, i.e. it assumes the energy to create $M$ excitations is exactly $M$ times the energy to create one excitation. This is not true for the model in Eq.~(\ref{eq:Hamil}) because the two-level systems are saturable. 
However, such effects were shown in~\cite{Cortese2017} to not
significantly change the behavior.  (We also consider this further below.)

The integrals over $\theta_i$ can be transformed to an integral
over the end-to-end distribution of a polymer.  Specifically, we
consider 
\begin{equation*}
\vec R \equiv \begin{pmatrix} R_x \\ R_y \end{pmatrix} = \sum_i \begin{pmatrix} \cos 2\theta_i \\ \sin 2\theta_i\end{pmatrix},
\end{equation*}
which is the vector formed by adding unit vectors each oriented at angle $2\theta_i$.
Then, using $\sum_i \cos^2\theta_i = \sum_i (1+\cos(2\theta_i))/2 = (N+R_x)/2$, the integral can be rewritten as
\begin{displaymath}
  \mathcal{Z} = \int d\vec{R} P_N(\vec{R})
  \exp\left( N a \sqrt{\frac{1+ R_x/N}{2}} \right)
\end{displaymath}
where $a=\beta (M/N) g\sqrt{N}$ and
$P_N(\vec{R})$ is the probability distribution of the vector $\vec{R}$, which can be considered as a polymer chain of $N$ links.  The peak of this probability distribution is at $\vec{R}=0$, corresponding to entirely disordered dipoles, and the variance of this distribution scales as  $\langle \vec{R}^2 \rangle  \propto N$  as expected for a random walk.

In writing the exponent above, we have explicitly separated the scaling with system size $N$, from the intensive quantity $a$ which depends on the  excitation \emph{density}
$M/N$ and the quantity $g\sqrt{N}$ which, as discussed above, remains finite in
the limit of large $N$.  As discussed in~\cite{Cortese2017}, for the record
polariton splitting of $g\sqrt{N} \simeq 0.5$eV, this quantity would at room
temperature correspond to $a \simeq 20 (M/N)$.
The length $R_x$ can also be used to evaluate an
order parameter for the orientational ordering, i.e.
\begin{displaymath}
  \langle \cos^2 \theta \rangle 
  = \frac{1 + x}{2} , \qquad
  x = \left< \frac{R_x}{N} \right>.
\end{displaymath}
To explore ordering, we are interested in how $x$ evolves with the parameter
$a$.

Under the assumption that one may replace $P_N(\vec{R})$ with its large $N$
approximation from the central limit theorem, $P_N(\vec{R}) \simeq \exp( -
R^2/N) / (\pi N)$, one could evaluate the partition function.  (However,
as discussed further below, this approximation has limited validity).
Due to the
large parameter $N$,  one may use a saddle point evaluation, leading the to
statement that $x$ is given by the minimum of $x^2 - a\sqrt{(1+x)/{2}}$,
given by:
\begin{equation}
  \label{eq:gauss-polymer}
  a = 4 x_0 \sqrt{2(1+x_0)}
\end{equation}
The solution of this equation increases from $x_0=0$ at $a=0$ to reach
$x_0=1$ when $a=8$.  By definition, $|x_0|<1$, and so this Gaussian polymer approximation predicts that at $a=8$, a second
order transition occurs to a fully ordered state~\cite{Cortese2017}.

\section{Mean-field theory of orientational ordering}
\label{sec:mean-field-theory}

In this section we discuss an alternate approach to finding the partition
function of Eq.~(\ref{eq:Hamil}): mean-field theory, as has been discussed many
times for variants of the Dicke
model~\cite{wang73,eastham01,Marchetti2007a,Cwik2014}.  As discussed above,
we consider the grand canonical ensemble, so the effective Hamiltonian
becomes $ H - \mu  M$.  In making comparison to Ref.~\onlinecite{Cortese2017},
we will tune the excitation density by adjusting $\mu$. Within mean-field
theory, there is a transition between a normal state, with zero photon
number, and a condensed state.  Mean-field theory proceeds by assuming a
coherent state $|\alpha\rangle$ for the photons, and performing a
variational minimization over the coherent field amplitude.  In the normal
state, the minimum occurs at $\alpha=0$, while for the condensed state, the
minimum occurs at finite $\alpha$.
Such an approach can be rigorously justified by evaluating a path integral
form of the partition function, and noting that in the limit $N\to \infty$,
a saddle point expression becomes exact~\cite{eastham01}.  Such a procedure
implies $\mathcal{Z}=\exp(-\beta F)$ where:
\begin{equation}
  F= \inf_\alpha \left[
    (\omega-\mu) |\alpha|^2
    - 
    N k_B T \ln \left( \text{Tr} e^{-\beta h} \right) \right],
\end{equation}
with $h$ being the Hamiltonian of a single dipole in the presence of the coherent field, $\alpha$:
\begin{displaymath}
  h = \frac{1}{2} 
  \begin{pmatrix}
    \omega_0-\mu & 2 g \cos\theta_i \alpha^\ast \\
    2 g \cos\theta_i \alpha & - (\omega_0 - \mu)
  \end{pmatrix}.
\end{displaymath}
The trace appearing in the partition function involves both a trace
over two by two matrices, as well as a trace over angular orientations.

One can rewrite the above in terms of only intensive quantities by noting
the photon density $|\alpha|^2$ scales with $N$ in the condensed
state~\cite{wang73}, and so writing $|\alpha|^2 = N \rho$.  One then finds
\begin{align}
  \label{eq:mft-f}
  \frac{F}{N} &= \inf_{\rho>0} 
  \left[
    (\omega-\mu) \rho 
    - k_B T \ln \mathcal{Z}_{\text{2LS}}
  \right],
  \\
  \label{eq:mft-z}
  \mathcal{Z}_{\text{2LS}}
  &=
  \int\!\! d\theta \;2 \cosh\left(\frac{\beta E(\theta)}{2}\right),
\end{align}
where we have used the two-level system energy:
\begin{displaymath}
  E(\theta) = \sqrt{(\omega_0 - \mu)^2 + 4 (g\sqrt{N})^2 \rho \cos^2 \theta}.
\end{displaymath}
Once $\rho$ is known, the angular orientation can be found as
\begin{equation}
  \label{eq:mft-order}
  \langle \cos^2 \theta \rangle
  =
  \frac{1}{\mathcal{Z}_{\text{2LS}}}
  \int\!\! d\theta \cos^2(\theta) 
  \;2 \cosh\left(\frac{\beta E(\theta)}{2}\right).
\end{equation}

\begin{figure}[htp]
  \centering
  \includegraphics[width=3.2in]{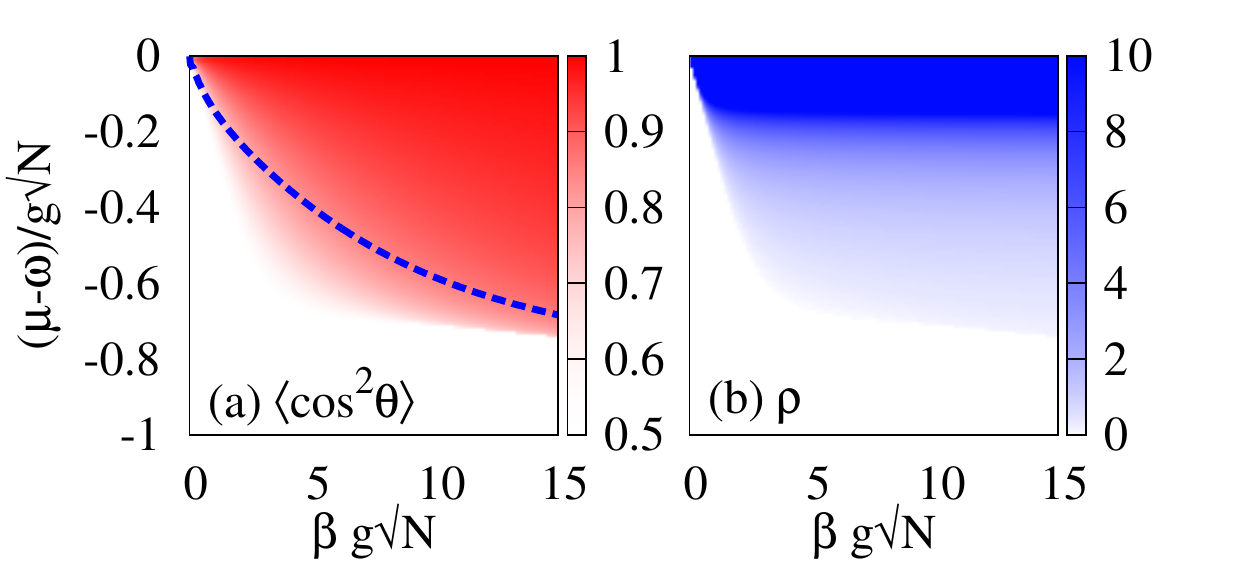}
  \caption{Mean-field theory results.  Panel (a) shows orientational
    ordering, $\langle \cos^2\theta \rangle$, while panel (b) shows the
    corresponding condensate density $\rho$.  The dashed line in panel (a) 
    corresponds to the prediction in Ref.~\onlinecite{Cortese2017} for 
    complete ordering, $\beta g\sqrt{N} (M/N) = 8$.  }
  \label{fig:mft}
\end{figure}
If we focus on the resonant case, $\omega=\omega_0$, the ordering
parameter depends on two dimensionless quantities, $\beta g\sqrt{N}$ and
$(\mu-\omega)/g\sqrt{N}$.  Figure~\ref{fig:mft} shows the orientational
ordering and condensate density as a function of these quantities.  When
the condensate density is zero, we see immediately from
Eq.~(\ref{eq:mft-order}) that $\langle \cos^2 \theta \rangle =0.5$, as
the energy $E(\theta)$ becomes independent of $\theta$ when $\rho=0$.
Inside the condensed region the orientational order grows and
approaches $1$.  However, crucially we see that it grows smoothly
without any sharp transition.

For direct comparison the results of \citet{Cortese2017}, Eq.~(\ref{eq:gauss-polymer}), we must extract
the total excitation number, by considering the derivative of free energy
with chemical potential:
\begin{equation}
  \label{eq:mft-exc}
  \frac{M}{N} 
  =
  \rho + 
  \frac{1}{\mathcal{Z}_{\text{2LS}}}
  \int\!\! d\theta 
  \left[1 + \frac{\mu}{E(\theta)}\right] \cosh\left(
    \frac{\beta E}{2} 
  \right).
\end{equation}
The trajectory at $a \equiv \beta (M/N) g\sqrt{N} = 8$ is marked
by the blue dashed line in Fig.~\ref{fig:mft}.  We see this does not
correspond to any sharp transition of the orientational ordering.

\section{Monte Carlo integration}
\label{sec:monte-carlo-integr}

Having seen that the mean-field approach predicts no complete orientational
ordering at any finite occupation or temperature, we next compare this to exact numerics
at finite $N$.  Specifically, we consider the problem as defined in
Eq.~(\ref{eq:exact-Z}), and the corresponding orientational ordering quantified by:
\begin{equation}
  \label{eq:exact-op}
  \langle \cos^2 \theta \rangle
  =
  \frac{1}{\mathcal{Z}} \prod_i \int \!\!d\theta_i 
  \overline{\cos^2\theta} 
  \exp\left(
    N a  \sqrt{\overline{\cos^2\theta}}
  \right).
\end{equation}
where we have denoted $\overline{\cos^2\theta} \equiv \sum_i
\cos^2\theta_i/N$.  This
expression may be evaluated directly by Monte Carlo integration.
Specifically, we sample configurations $\{ \theta_i \}$, and
evaluate the expectation of the order parameter $\overline{\cos^2\theta}$
weighted by the Boltzmann factor $P_{\text{Boltz}} = \exp\left( N a
  \sqrt{\overline{\cos^2\theta} }\right)$.  To sample this efficiently, we
draw samples from a Gaussian approximation of the Boltzmann distribution,
i.e. $P_{\text{draw}}(\{ \theta_i \}) = \prod_i \exp\left( - a \theta_i^2/2
\right)$, and weight samples by the ratio
$P_{\text{Boltz}}/P_{\text{draw}}$.  The distribution $P_{\text{draw}}$ is
factorizable, hence it is easy to draw samples from this distribution.  In
addition $P_{\text{Boltz}} \simeq P_{\text{draw}}$ for small angles; at low
temperature only small angles are probable, so the sampling becomes
efficient in this limit.  We may also note that in this limit, with independent
$\theta_i$, this problem is self averaging, so the sampling error
reduces at large $N$.

\begin{figure}[htp]
  \centering
  \includegraphics[width=3.2in]{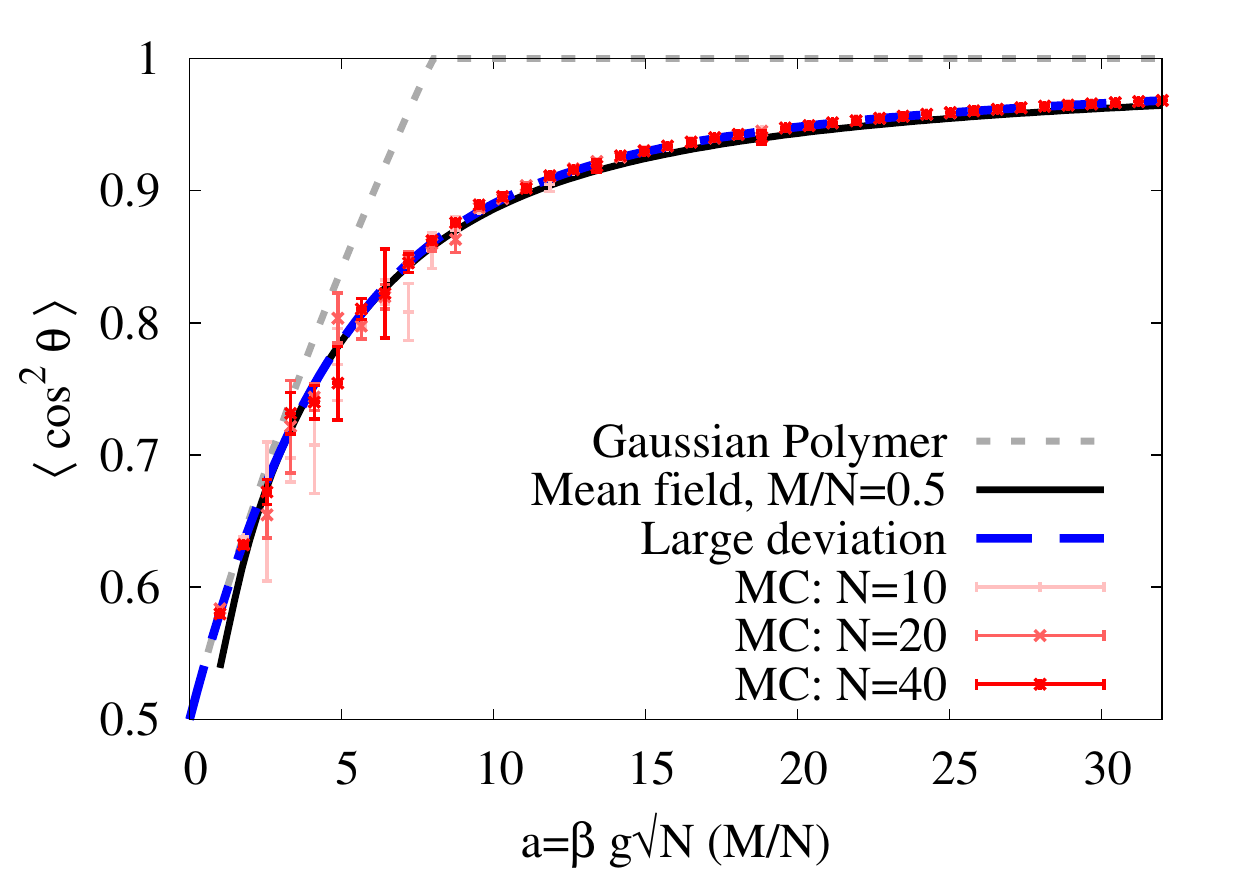}
  \caption{Order parameter as a function of $a=\beta g\sqrt{N} (M/N)$,
    comparing Monte Carlo results to mean-field and Gaussian polymer
    results.  Monte Carlo results are shown as points, with error bars
    reflecting the sampling error, with each point using
    20000 samples.  The gray short dashed line indicates the
    Gaussian Polymer prediction~\cite{Cortese2017}, $(1+x_0)/2$ with $x_0$ given by
    Eq.~(\ref{eq:gauss-polymer}).  The black line shows mean-field prediction
    in the low temperature, low excitation limit as discussed in the text.  The
    blue long dashed line shows the large deviation result.}
  \label{fig:mc-mf}
\end{figure}

The order parameter calculated by this Monte Carlo approach is shown in
Fig.~\ref{fig:mc-mf}, for various values of $N$, in each case choosing
excitation fraction $M/N=0.5$.  We clearly see that while the results have
converged with respect to $N$ (indeed, even $N=10$ seems converged), they
do not converge on the result of the polymer model described in 
Sec.~\ref{sec:model-prev-results}.
.

In order to compare these exact numerics to mean-field theory, 
we must make a number of modifications to the mean-field equations.  These  follow from the fact that Eq.~(\ref{eq:exact-Z})
and the results following from it (a) assume a thermal population of only the lower polariton mode, and (b) neglect saturation effects arising from the non-linearity of two-level systems.  
To address point (a), we restrict to the lower energy branch,
 modifying the mean-field theory by writing $
\mathcal{Z}_{\text{2LS}} = \int\!\! d\theta e^{\beta E(\theta)/2}$ in
contrast to Eq.~(\ref{eq:mft-z}). Such a replacement is valid if $\beta g
\sqrt{N} \gg 1$.  
To address point (b), we work at low excitation
fraction $M/N=0.5$. In the limit $M/N \to 0$, the excitation of each two-level system is small, and so the saturation at a maximum of one excitation per dipole has little effect.   One may note that the low temperature
and low excitation limits are
consistent: if $M/N \ll 1$, then $\beta g \sqrt{N} \gg 1$ for all non-zero
$a$.  To fix the excitation fraction $M/N$, we use
Eq.~(\ref{eq:mft-exc}) with the replacement $\cosh(\beta E/2) \to \exp(\beta E/2)$ as a self consistent equation to fix $\mu$.

\section{Large deviations}
\label{sec:large-devi-form}

We next turn to consider whether the Polymer model discussed in
\citet{Cortese2017} can be improved to match the behavior seen
from the above Monte Carlo results.  The
approximation which leads to the mismatch is the replacing of
$P_N(\vec{R})$ by a Gaussian distribution.  The reason this
approximation fails can be understood as follows: The Gaussian
distribution is valid for ``typical'' values of $\vec{R}$, which,
means $|\vec{R}| \simeq \mathcal{O}(\sqrt{N})$.
However, in the limit of large $a$, the matter-light coupling biases one
towards atypical configurations, where $R_x \simeq \mathcal{O}(N)$.  Such
values are deep in the tail of the probability distribution; they
correspond to large deviations from the mean, and are not given by the
Gaussian approximation.  In fact, in the limit $N \to \infty$, any non-zero
value of $x = R_x/N$ corresponds to a large deviation.

\subsection{Analytic large deviation formulation}
\label{sec:anal-large-devi-form}

Fortunately, there is simple approach to extract the probability of
large deviations, as reviewed, e.g., by \citet{Touchette2009}.
We are interested in finding the probability $P_N(x=R_x/N)$ and so
we use the standard results of the large deviation formulation:
\begin{equation}
  \label{eq:12}
  P_N(x) \simeq e^{ - N w (x)}, \quad
  w(x) = \sup_{s} \left[x s - \lambda(s) \right], 
\end{equation}
where $\lambda(s)$ is the generating function at large $N$:
\begin{equation}
  \lambda(s) = \lim_{N \to \infty}\left[
    \frac{1}{N} \ln \langle e^{ s R_x} \rangle \right].
\end{equation}
This can be directly evaluated for a model of an $N$ link polymer chain, 
\begin{displaymath}
  \langle e^{ s R_x} \rangle =
  \prod_i \int\!\!d\theta_i \exp\left[ s \sum_i\cos(\theta_i) \right]
  =
  \left[ I_0(s) \right]^N,
\end{displaymath}
where $I_0(s)$ is the modified Bessel function of the first kind.
Thus, we have:
\begin{equation}
  \label{eq:ld-wx}
  w(x) = \sup_{s} \left[x s - \ln(I_0(s)) \right].
\end{equation}
This function $w(x)$ replaces the quadratic exponent in the Gaussian
polymer approximation. We can then use this to find an alternative to
Eq.~(\ref{eq:gauss-polymer}) for determining $x_0$.  In terms of
$x$, the partition function can be written as:
\begin{displaymath}
  \mathcal{Z} \propto \int dx
  \exp\left( -N w(x)
    + N a \sqrt{\frac{1 + x}{2}}
  \right),
\end{displaymath}
and it is clear that at large $N$, this can be approximated by its saddle
point, $x_0$,  given by solving:
\begin{displaymath}
  \left.\frac{d w}{dx}\right|_{x_0} = \frac{a}{2\sqrt{2 (1+x_0)}}.
\end{displaymath}
To evaluate $w(x)$, we note that the supremum over $s$ in Eq.~(\ref{eq:ld-wx})
is solved by  $s_0(x)$ such that:
\begin{displaymath}
  x = \left.\frac{d}{ds} \ln I_0(s)\right|_{s=s_0(x)} 
  = \frac{I_1(s_0(x))}{I_0(s_0(x))}.
\end{displaymath}
We may then use this to evaluate the derivative of $w(x)$, writing:
\begin{displaymath}
  \frac{d w(x)}{dx} 
  = 
  \frac{\partial w(x)}{\partial x}
  +
  \frac{\partial w(x) }{\partial s_0} \frac{ d s_0}{d x}
  =
  s_0(x).  
\end{displaymath}
In the final expression we used the explicit form of $w(x)$ to
evaluate the first term, and the fact that the second
term  vanishes by the definition of $s_0$.
Putting these  together we find that $x_0$ is determined by the 
pair of equations:
\begin{equation}
  \label{eq:8}
  s_0 = \frac{a}{2\sqrt{2 (1+x_0)}}, \qquad
  x_0 = \frac{I_1(s_0)}{I_0{(s_0)}}.
\end{equation}
Solving these simultaneous equations gives the blue long dashed line in
Fig.~\ref{fig:mc-mf} which almost perfectly matches the Monte Carlo result.
One may also explicitly see this expression predicts perfect orientation
only at zero temperature,  i.e., as $a \to \infty$. Assuming $x_0
\simeq 1$ one finds $s_0 \simeq a/4$, giving an explicit expression for
$x_0$ which approaches one asymptotically from below.

\subsection{Recovering large deviation result from mean-field theory}
\label{sec:recov-large-devi}

The mean-field theory results at low excitation and low
temperature appear to match both the large deviation analytic form
and the Monte Carlo results well.  Here we show that this match
can in fact be seen analytically, by considering the mean-field
equations perturbatively in the limit $M/N \to 0$.  Crucially, since
the expression for $M/N$ in Eq.~(\ref{eq:mft-exc}) contains a term
$\rho$, the limit of low density requires that we consider $\rho$
small.  In this limit we may expand
\begin{displaymath}
  E(\theta) \simeq 
  |\omega_0-\mu| + \frac{2 \rho g^2 N \cos^2\theta}{|\omega_0-\mu|}.
\end{displaymath}
Using this expansion, combined with the restriction to the lower
branch in evaluating $\mathcal{Z}_{\text{2LS}}$, we find that the density
equation becomes:
\begin{align}
  \label{eq:mft-exc-limit}
  \frac{M}{N} 
  &=
  \rho + 
  \frac{1}{\mathcal{Z}_{\text{2LS}}}
  \int\!\! d\theta 
  \frac{g^2N\rho \cos^2(\theta)}{|\omega_0-\mu|^2} 
  e^{\beta E(\theta)/2}
  \nonumber\\&=
  \rho\left(
    1+ \frac{g^2N \langle \cos^2(\theta) \rangle}{|\omega_0-\mu|^2}  
  \right)    
\end{align}
The angular average also simplifies, as we can write
\begin{displaymath}
  E(\theta) \simeq E_0 + E_1 \cos(2 \theta)
\end{displaymath}
which allows angular integrals to be rewritten in terms of modified Bessel
functions, namely
\begin{equation}
  \label{eq:mft-angle}
  \langle \cos^2 \theta \rangle
  =
  \frac{1+x}{2}, \qquad
  x=\frac{I_1(\beta E_1/2)}{I_0(\beta E_1/2)}
\end{equation}
where $E_1={\rho g^2 N }/{|\omega_0-\mu|}$.
We can then combine this with the self-consistency condition, from
evaluating the infinum in Eq.~(\ref{eq:mft-f}) which gives:
\begin{align}
  (\omega-\mu) &=
  \frac{1}{\mathcal{Z}_{\text{2LS}}}
  \int\!\! d\theta 
  \frac{1}{2}
  \frac{d E(\theta)}{d\rho}
  e^{\beta E(\theta)/2}
  \nonumber\\&\simeq
  \frac{ g^2N \langle\cos^2(\theta)\rangle}{|\omega_0-\mu|} 
\end{align}
In the resonant limit $\omega=\omega_0$, assuming that $\omega>\mu$ as is
required for physical solutions, we then find that $\omega-\mu = g
\sqrt{N} \sqrt{\langle\cos^2(\theta)\rangle}$, and thus $\rho = M/2N$. 
Inserting this into the definition of $E_1$ we find:
\begin{equation}
  \label{eq:mft-sc}
  \frac{\beta E_1}{2} = 
  \frac{\beta (M/N) g \sqrt{N}}{2\sqrt{2(1+x)}}
  = \frac{a}{2\sqrt{2(1+x)}}
\end{equation}
Together, Eq.~(\ref{eq:mft-angle},\ref{eq:mft-sc}) precisely recover
the large deviation result, hence the agreement of mean-field
theory in this limit.

\section{Saturation effects }
\label{sec:multiple-excitation}

As noted earlier, the polymer model and Monte Carlo results above use the approximation that the energy of an $M$ polariton state, $\epsilon_M$ is equal to $M$ times the one polariton state, $\epsilon_M \simeq M \epsilon_1$.  Such an assumption is incorrect for Eq.~(\ref{eq:Hamil}), as this model is not linear --- it involves saturable two-level systems.   In this section we discuss how our results change when we take this saturation and non-linearity into account.  

\begin{figure}[htp]
  \centering
  \includegraphics[width=3.2in]{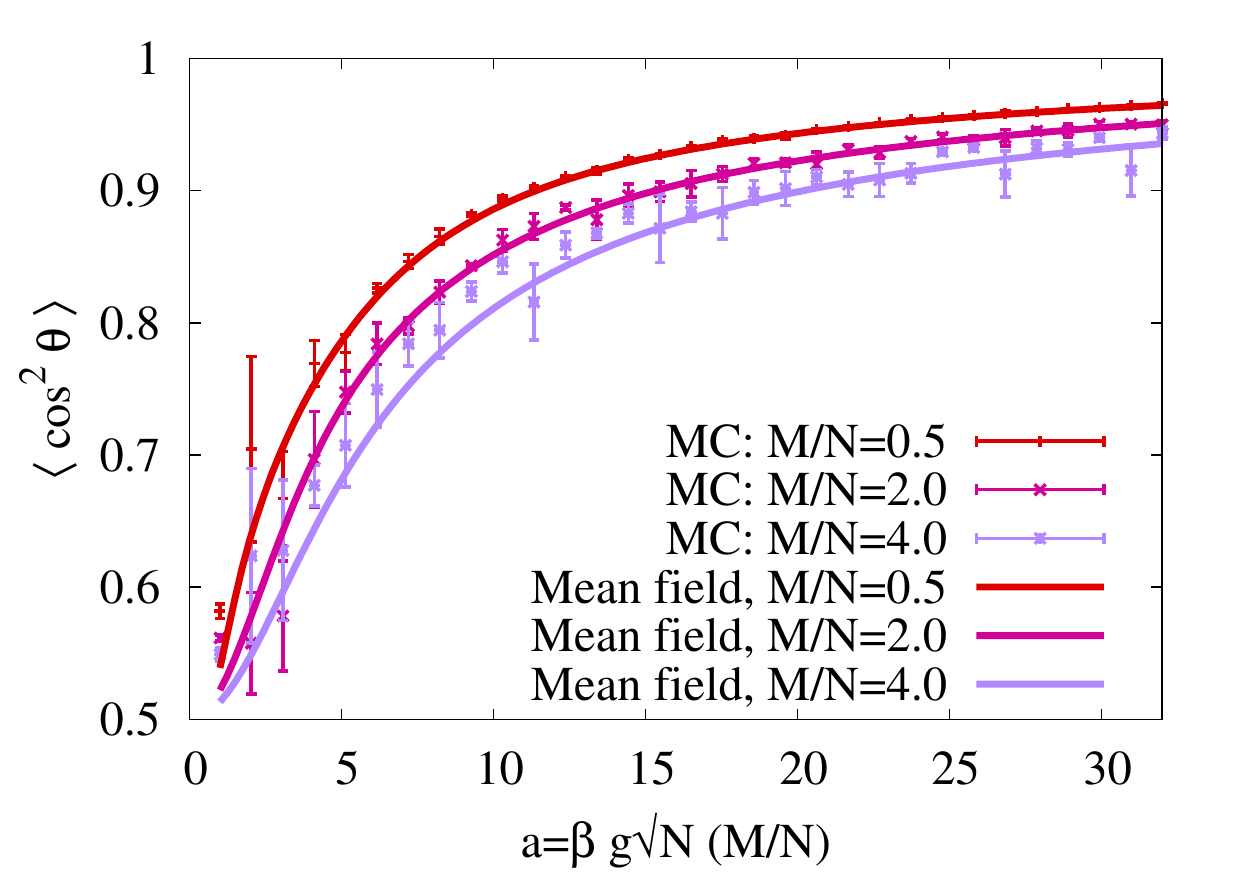}
  \caption{Comparison of mean-field theory results at various excitation
    fractions to Monte Carlo calculations allowing for saturation effects.
    Excitation fraction increases from top ($M/N=0.5$) to bottom ($M/N=4.0$).
    The Monte Carlo results correspond to  $N=10$ molecules; 
    each point is calculated with $4000$ samples. }
  \label{fig:mc-mf-saturation}
\end{figure}

In contrast to the Monte Carlo results and polymer model, the mean-field approach makes no assumption of linearity.  i.e., the mean-field theory is based on solving the exact energies of two-level atoms in the presence of a coherent field.  Thus, for the mean-field approach we can directly determine the effect of saturation by considering the behavior at different filling fractions $M/N$.   This is shown in the solid lines in
Fig.~\ref{fig:mc-mf-saturation} which show the mean-field results for the orientational ordering.  Each line corresponds to a different excitation fraction, and the horizontal axis is the variable $a=\beta g \sqrt{N} (M/N)$, which we may still tune by adjusting $\beta g \sqrt{N}$.   We see that the lines do not fall on top of each other, indicating that the results depend on the values $M/N$ and $\beta g\sqrt{N}$ separately --- there is no reduction to a single result depending only on $a=\beta g\sqrt{N} (M/N)$.  This indicates an effect of saturation, as it means we no longer can match the large deviation result, as we could in the limit $M/N \to 0$.

Next, we consider how to modify the Monte Carlo integration to recover correct results at a finite excitation density.  As noted above, the error in the polymer model comes by assuming $\epsilon_M(\{\theta_i\}) \simeq M \epsilon_1(\{\theta_i\} \equiv -M g \sqrt{\sum_i \cos^2\theta_i}$.  To correct
this, we must therefore replace Eq.~(\ref{eq:exact-Z}) with
\begin{equation}
  \label{eq:exact-Z-nl}
  \mathcal{Z} = \prod_i \int \!\!d\theta_i 
  \exp\left( - \beta \epsilon_M(\{\theta_i\})\right),
\end{equation}
using the true energy of the $M$ excitation state of Eq.~(\ref{eq:Hamil}).
We find this energy numerically: For each configuration of angles $\{\theta_i\}$
we construct the  Hamiltonian in the $M$ excitation subspace, find its
lowest eigenvalue, and use this value as $\epsilon_M(\{\theta_i\})$.
For $M \geq N$, this requires us to find the lowest eigenvalue of a $2^N$ dimensional matrix for each configuration $\{ \theta_i \}$; when
$M< N$ the matrix size can be smaller.  For $N=10$, we thus require
eigenvalues of a $1024 \times 1024$ matrix for each configuration, this
is achievable, but computationally costly so we take only $4000$ samples. The results of this are also shown as the data points in
Fig.~\ref{fig:mc-mf-saturation}.  There is reasonable agreement between
the mean-field and $N=10$ Monte Carlo results for $M/N=0.5$ and $M/N=2.0$; the
agreement is less clear at the highest excitation level; this is likely
due to the finite size effects for $N=10$.

\section{Summary}
\label{sec:summary}

In this article we have shown that the evolution of orientational
order with temperature and density  can be captured
both through a large deviation formula, and through mean-field theory.
The large deviation approach is derived from an approximate partition function valid 
in the low excitation density limit.  In this limit mean-field theory exactly
reproduces the large deviation approach.  Furthermore, we have shown that
away from this limit, mean-field theory  matches  exact numerics well,
indicating the validity of mean-field theory for general excitation
densities.  The behavior we find shows a  smooth evolution of ordering
with excitation and temperature, and does not undergo any sharp transition
to a fully ordered state. 

An important conclusion of this work is that mean-field theory can indeed
be used as a simple and adaptable theoretical tool to understand a variety
of other related models. i.e., one may replace rotational orientation with
a variety of ways of dressing the Dicke model such as deformation of a
molecule or vibrational state etc. The case of vibrational dressing using
this mean-field approach was already considered in
Ref.~\onlinecite{Cwik2014}.

The validity of mean-field theory for such problems is also useful in that 
mean-field approaches can be easily adapted to  non-equilibrium situations.
An extension to the non-equilibrium version of this problem would
be an interesting challenge for future work, exploring how incoherent
excitation balanced with cavity loss can potentially lead to a modification
of orientational ordering.  Another related extension involves considering
multiple polarizations of light, and the relation between orientational
order and the polarization of the condensate.  This can potentially
form a ``strong coupling'' analogue to recent discussions of the polarization
state in a weak coupling photon BEC~\cite{Moodie2017,Greveling2017}.

\begin{acknowledgments}
  We are grateful to S. De Liberato for useful discussions.
  JK and PGK acknowledges financial support from EPSRC program ``Hybrid
  Polaritonics'' (EP/M025330/1).  
\end{acknowledgments}

\begin{appendix}

\section{Three dimensional orientation}
\label{sec:large-devi-form-1}

The approach outlined above allows simple extensions to other models.
For example, we can consider dipoles allowed to rotate in three dimensions,
by considering
\begin{equation}
  \label{eq:exact-Z-3D}
  \mathcal{Z} = \prod_i \int \!\!d\theta_i \sin(\theta_i) \int \!\! d\phi_i
  \exp\left( \beta M g \sqrt{\sum_i \cos^2\theta_i}\right).
\end{equation}
The $\phi$ integral is of course trivial here (as we have chosen the
electric field to be aligned along the $z$ axis).  The $\theta$ integral
is modified by the changed integral measure.

\begin{figure}[htpb]
  \centering
  \includegraphics[width=3.2in]{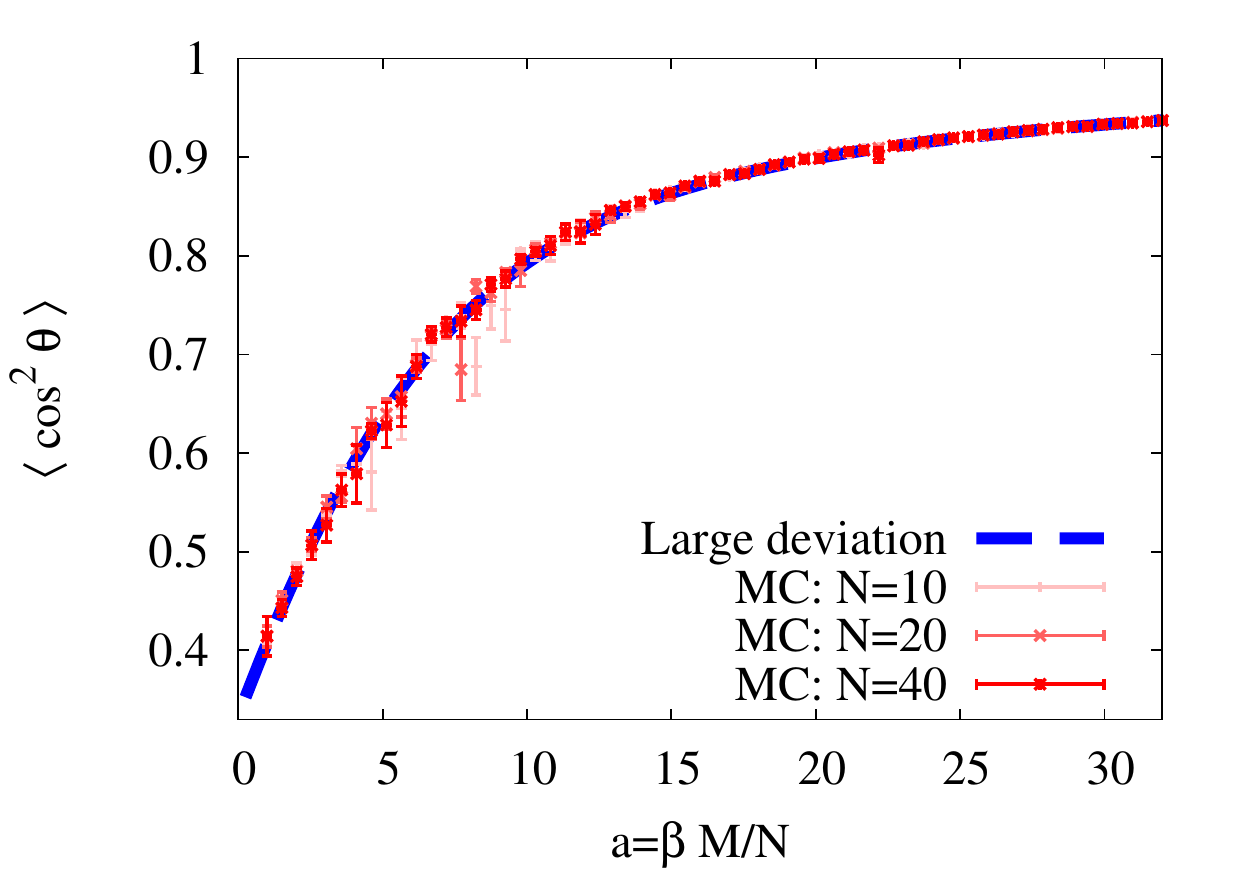}
  \caption{Comparison of Monte Carlo using the one-excitation state, and
    large deviation results for rotational orientation of three dimensional
    dipoles. Lines are as in Fig.~\ref{fig:mc-mf}; data correspond
    to 2000 samples for each point.}
  \label{fig:mc-mft-3d}
\end{figure}

The large deviation approach remains applicable, in terms of the
variable $x = \sum_i \cos(2\theta_i)/N$. The generating
function $\lambda(s)$ now takes a different form, namely:
\begin{displaymath}
  \lambda(s) = - s - \frac{1}{2} \ln(s)
  +
  \ln \left[ \erfi(\sqrt{2s}) \right] + \text{const.}.
\end{displaymath}
From this expression, we then find the self consistent equations for $x_0,
s_0$ take the form:
\begin{align}
  s_0 &= \frac{a}{2\sqrt{2 (1+x_0)}}, \\
  x_0 &= -1 - \frac{1}{2s_0} 
  + \sqrt{\frac{2}{\pi s_0}} \frac{e^{2 s_0}}{\erfi(\sqrt{2s_0})}.
\end{align}
Figure~\ref{fig:mc-mft-3d} shows these results, again comparing to Monte
Carlo integration of Eq.~(\ref{eq:exact-Z-3D}).  Once again, at large
$a$ the results asymptotically approach complete alignment, but without any
sharp transition.  The behavior at small $a$ differs from the previous
case: at infinite temperatures, the angular average now gives $\langle \cos^2\theta \rangle = 1/3$ rather than $1/2$.

As well as the large deviation formula, mean-field theory can also be directly
applied to this problem.  This corresponds to replacing the integral
over angle $\theta$ in Eq.~(\ref{eq:mft-z}) by:
\begin{displaymath}
  \mathcal{Z}_{\text{2LS}}
  =
  \int d \theta \sin(\theta) \int d\phi
  2 \cosh\left( \frac{\beta E(\theta)}{2} \right).
\end{displaymath}
It is once again possible to show that the mean-field result, in the 
limits $M/N \to 0$ and $\beta g\sqrt{N} \gg 1$, recovers the same form
as the large deviation expression.

\end{appendix}


\begin{thebibliography}{55}%
\makeatletter
\providecommand \@ifxundefined [1]{%
 \@ifx{#1\undefined}
}%
\providecommand \@ifnum [1]{%
 \ifnum #1\expandafter \@firstoftwo
 \else \expandafter \@secondoftwo
 \fi
}%
\providecommand \@ifx [1]{%
 \ifx #1\expandafter \@firstoftwo
 \else \expandafter \@secondoftwo
 \fi
}%
\providecommand \natexlab [1]{#1}%
\providecommand \enquote  [1]{``#1''}%
\providecommand \bibnamefont  [1]{#1}%
\providecommand \bibfnamefont [1]{#1}%
\providecommand \citenamefont [1]{#1}%
\providecommand \href@noop [0]{\@secondoftwo}%
\providecommand \href [0]{\begingroup \@sanitize@url \@href}%
\providecommand \@href[1]{\@@startlink{#1}\@@href}%
\providecommand \@@href[1]{\endgroup#1\@@endlink}%
\providecommand \@sanitize@url [0]{\catcode `\\12\catcode `\$12\catcode
  `\&12\catcode `\#12\catcode `\^12\catcode `\_12\catcode `\%12\relax}%
\providecommand \@@startlink[1]{}%
\providecommand \@@endlink[0]{}%
\providecommand \url  [0]{\begingroup\@sanitize@url \@url }%
\providecommand \@url [1]{\endgroup\@href {#1}{\urlprefix }}%
\providecommand \urlprefix  [0]{URL }%
\providecommand \Eprint [0]{\href }%
\providecommand \doibase [0]{http://dx.doi.org/}%
\providecommand \selectlanguage [0]{\@gobble}%
\providecommand \bibinfo  [0]{\@secondoftwo}%
\providecommand \bibfield  [0]{\@secondoftwo}%
\providecommand \translation [1]{[#1]}%
\providecommand \BibitemOpen [0]{}%
\providecommand \bibitemStop [0]{}%
\providecommand \bibitemNoStop [0]{.\EOS\space}%
\providecommand \EOS [0]{\spacefactor3000\relax}%
\providecommand \BibitemShut  [1]{\csname bibitem#1\endcsname}%
\let\auto@bib@innerbib\@empty
\bibitem [{\citenamefont {Agranovich}(1968)}]{AgranovichRussian}%
  \BibitemOpen
  \bibfield  {author} {\bibinfo {author} {\bibfnamefont {V.~M.}\ \bibnamefont
  {Agranovich}},\ }\href@noop {} {\emph {\bibinfo {title} {{The Theory of
  Excitons}}}}\ (\bibinfo  {publisher} {Nauka},\ \bibinfo {address} {Moscow},\
  \bibinfo {year} {1968})\BibitemShut {NoStop}%
\bibitem [{\citenamefont {Agranovich}(2009)}]{Agranovich2009a}%
  \BibitemOpen
  \bibfield  {author} {\bibinfo {author} {\bibfnamefont {V.~M.}\ \bibnamefont
  {Agranovich}},\ }\href@noop {} {\emph {\bibinfo {title} {{Excitations in
  Organic Solids}}}}\ (\bibinfo  {publisher} {Oxford University Press},\
  \bibinfo {address} {Oxford},\ \bibinfo {year} {2009})\BibitemShut {NoStop}%
\bibitem [{\citenamefont {Barford}(2013)}]{barford13}%
  \BibitemOpen
  \bibfield  {author} {\bibinfo {author} {\bibfnamefont {W.}~\bibnamefont
  {Barford}},\ }\href@noop {} {\emph {\bibinfo {title} {{Electronic and optical
  properties of conjugated polymers}}}}\ (\bibinfo  {publisher} {Oxford
  University Press},\ \bibinfo {address} {Oxford},\ \bibinfo {year}
  {2013})\BibitemShut {NoStop}%
\bibitem [{\citenamefont {Schwartz}\ \emph {et~al.}(2011)\citenamefont
  {Schwartz}, \citenamefont {Hutchison}, \citenamefont {Genet},\ and\
  \citenamefont {Ebbesen}}]{Schwartz11}%
  \BibitemOpen
  \bibfield  {author} {\bibinfo {author} {\bibfnamefont {T.}~\bibnamefont
  {Schwartz}}, \bibinfo {author} {\bibfnamefont {J.~A.}\ \bibnamefont
  {Hutchison}}, \bibinfo {author} {\bibfnamefont {C.}~\bibnamefont {Genet}}, \
  and\ \bibinfo {author} {\bibfnamefont {T.~W.}\ \bibnamefont {Ebbesen}},\
  }\href {\doibase 10.1103/PhysRevLett.106.196405} {\bibfield  {journal}
  {\bibinfo  {journal} {Phys. Rev. Lett.}\ }\textbf {\bibinfo {volume} {106}},\
  \bibinfo {pages} {196405} (\bibinfo {year} {2011})}\BibitemShut {NoStop}%
\bibitem [{\citenamefont {Hutchison}\ \emph {et~al.}(2012)\citenamefont
  {Hutchison}, \citenamefont {Schwartz}, \citenamefont {Genet}, \citenamefont
  {Devaux},\ and\ \citenamefont {Ebbesen}}]{Hutchison12}%
  \BibitemOpen
  \bibfield  {author} {\bibinfo {author} {\bibfnamefont {J.~A.}\ \bibnamefont
  {Hutchison}}, \bibinfo {author} {\bibfnamefont {T.}~\bibnamefont {Schwartz}},
  \bibinfo {author} {\bibfnamefont {C.}~\bibnamefont {Genet}}, \bibinfo
  {author} {\bibfnamefont {E.}~\bibnamefont {Devaux}}, \ and\ \bibinfo {author}
  {\bibfnamefont {T.~W.}\ \bibnamefont {Ebbesen}},\ }\href@noop {} {\bibfield
  {journal} {\bibinfo  {journal} {Ang. Chem. Int. Ed.}\ }\textbf {\bibinfo
  {volume} {51}},\ \bibinfo {pages} {1592} (\bibinfo {year}
  {2012})}\BibitemShut {NoStop}%
\bibitem [{\citenamefont {Herrera}\ and\ \citenamefont
  {Spano}(2016)}]{Herrera15}%
  \BibitemOpen
  \bibfield  {author} {\bibinfo {author} {\bibfnamefont {F.}~\bibnamefont
  {Herrera}}\ and\ \bibinfo {author} {\bibfnamefont {F.~C.}\ \bibnamefont
  {Spano}},\ }\href {\doibase 10.1103/PhysRevLett.116.238301} {\bibfield
  {journal} {\bibinfo  {journal} {Phys. Rev. Lett.}\ }\textbf {\bibinfo
  {volume} {116}},\ \bibinfo {pages} {238301} (\bibinfo {year}
  {2016})}\BibitemShut {NoStop}%
\bibitem [{\citenamefont {Galego}\ \emph {et~al.}(2016)\citenamefont {Galego},
  \citenamefont {Garcia-Vidal},\ and\ \citenamefont {Feist}}]{Galego2016}%
  \BibitemOpen
  \bibfield  {author} {\bibinfo {author} {\bibfnamefont {J.}~\bibnamefont
  {Galego}}, \bibinfo {author} {\bibfnamefont {F.~J.}\ \bibnamefont
  {Garcia-Vidal}}, \ and\ \bibinfo {author} {\bibfnamefont {J.}~\bibnamefont
  {Feist}},\ }\href {\doibase 10.1038/ncomms13841} {\bibfield  {journal}
  {\bibinfo  {journal} {Nat. Commun.}\ }\textbf {\bibinfo {volume} {7}},\
  \bibinfo {pages} {13841} (\bibinfo {year} {2016})}\BibitemShut {NoStop}%
\bibitem [{\citenamefont {Kowalewski}\ \emph {et~al.}(2016)\citenamefont
  {Kowalewski}, \citenamefont {Bennett},\ and\ \citenamefont
  {Mukamel}}]{Kowalewski2016}%
  \BibitemOpen
  \bibfield  {author} {\bibinfo {author} {\bibfnamefont {M.}~\bibnamefont
  {Kowalewski}}, \bibinfo {author} {\bibfnamefont {K.}~\bibnamefont {Bennett}},
  \ and\ \bibinfo {author} {\bibfnamefont {S.}~\bibnamefont {Mukamel}},\ }\href
  {\doibase 10.1021/acs.jpclett.6b00864} {\bibfield  {journal} {\bibinfo
  {journal} {J. Phys. Chem. Lett.}\ }\textbf {\bibinfo {volume} {7}},\ \bibinfo
  {pages} {2050} (\bibinfo {year} {2016})}\BibitemShut {NoStop}%
\bibitem [{\citenamefont {Flick}\ \emph
  {et~al.}(2017{\natexlab{a}})\citenamefont {Flick}, \citenamefont
  {Ruggenthaler}, \citenamefont {Appel},\ and\ \citenamefont
  {Rubio}}]{Flick2017a}%
  \BibitemOpen
  \bibfield  {author} {\bibinfo {author} {\bibfnamefont {J.}~\bibnamefont
  {Flick}}, \bibinfo {author} {\bibfnamefont {M.}~\bibnamefont {Ruggenthaler}},
  \bibinfo {author} {\bibfnamefont {H.}~\bibnamefont {Appel}}, \ and\ \bibinfo
  {author} {\bibfnamefont {A.}~\bibnamefont {Rubio}},\ }\href {\doibase
  10.1073/pnas.1615509114} {\bibfield  {journal} {\bibinfo  {journal} {Proc.
  Natl. Acad. Sci.}\ }\textbf {\bibinfo {volume} {114}},\ \bibinfo {pages}
  {3026} (\bibinfo {year} {2017}{\natexlab{a}})}\BibitemShut {NoStop}%
\bibitem [{\citenamefont {Galego}\ \emph {et~al.}(2017)\citenamefont {Galego},
  \citenamefont {Garcia-Vidal},\ and\ \citenamefont {Feist}}]{galego17}%
  \BibitemOpen
  \bibfield  {author} {\bibinfo {author} {\bibfnamefont {J.}~\bibnamefont
  {Galego}}, \bibinfo {author} {\bibfnamefont {F.~J.}\ \bibnamefont
  {Garcia-Vidal}}, \ and\ \bibinfo {author} {\bibfnamefont {J.}~\bibnamefont
  {Feist}},\ }\href {\doibase 10.1103/PhysRevLett.119.136001} {\enquote
  {\bibinfo {title} {{Many-Molecule Reaction Triggered by a Single Photon in
  Polaritonic Chemistry}},}\ } (\bibinfo {year} {2017})\BibitemShut {NoStop}%
\bibitem [{\citenamefont {Orgiu}\ \emph {et~al.}(2015)\citenamefont {Orgiu},
  \citenamefont {George}, \citenamefont {Hutchison}, \citenamefont {Devaux},
  \citenamefont {Dayen}, \citenamefont {Doudin}, \citenamefont {Stellacci},
  \citenamefont {Genet}, \citenamefont {Schachenmayer}, \citenamefont {Genes},
  \citenamefont {Pupillo}, \citenamefont {Samor{\`{\i}}},\ and\ \citenamefont
  {Ebbesen}}]{orgiu15}%
  \BibitemOpen
  \bibfield  {author} {\bibinfo {author} {\bibfnamefont {E.}~\bibnamefont
  {Orgiu}}, \bibinfo {author} {\bibfnamefont {J.}~\bibnamefont {George}},
  \bibinfo {author} {\bibfnamefont {J.~A.}\ \bibnamefont {Hutchison}}, \bibinfo
  {author} {\bibfnamefont {E.}~\bibnamefont {Devaux}}, \bibinfo {author}
  {\bibfnamefont {J.~F.}\ \bibnamefont {Dayen}}, \bibinfo {author}
  {\bibfnamefont {B.}~\bibnamefont {Doudin}}, \bibinfo {author} {\bibfnamefont
  {F.}~\bibnamefont {Stellacci}}, \bibinfo {author} {\bibfnamefont
  {C.}~\bibnamefont {Genet}}, \bibinfo {author} {\bibfnamefont
  {J.}~\bibnamefont {Schachenmayer}}, \bibinfo {author} {\bibfnamefont
  {C.}~\bibnamefont {Genes}}, \bibinfo {author} {\bibfnamefont
  {G.}~\bibnamefont {Pupillo}}, \bibinfo {author} {\bibfnamefont
  {P.}~\bibnamefont {Samor{\`{\i}}}}, \ and\ \bibinfo {author} {\bibfnamefont
  {T.~W.}\ \bibnamefont {Ebbesen}},\ }\href {\doibase 10.1038/nmat4392}
  {\bibfield  {journal} {\bibinfo  {journal} {Nat. Mater.}\ }\textbf {\bibinfo
  {volume} {14}},\ \bibinfo {pages} {1123} (\bibinfo {year}
  {2015})}\BibitemShut {NoStop}%
\bibitem [{\citenamefont {Feist}\ and\ \citenamefont
  {Garcia-Vidal}(2015)}]{Feist2014}%
  \BibitemOpen
  \bibfield  {author} {\bibinfo {author} {\bibfnamefont {J.}~\bibnamefont
  {Feist}}\ and\ \bibinfo {author} {\bibfnamefont {F.~J.}\ \bibnamefont
  {Garcia-Vidal}},\ }\href {\doibase 10.1103/PhysRevLett.114.196402} {\bibfield
   {journal} {\bibinfo  {journal} {Phys. Rev. Lett.}\ }\textbf {\bibinfo
  {volume} {114}},\ \bibinfo {pages} {196402} (\bibinfo {year}
  {2015})}\BibitemShut {NoStop}%
\bibitem [{\citenamefont {Hagenm{\"{u}}ller}\ \emph {et~al.}(2018)\citenamefont
  {Hagenm{\"{u}}ller}, \citenamefont {Sch{\"{u}}tz}, \citenamefont
  {Schachenmayer}, \citenamefont {Genes},\ and\ \citenamefont
  {Pupillo}}]{Hagenmuller2018}%
  \BibitemOpen
  \bibfield  {author} {\bibinfo {author} {\bibfnamefont {D.}~\bibnamefont
  {Hagenm{\"{u}}ller}}, \bibinfo {author} {\bibfnamefont {S.}~\bibnamefont
  {Sch{\"{u}}tz}}, \bibinfo {author} {\bibfnamefont {J.}~\bibnamefont
  {Schachenmayer}}, \bibinfo {author} {\bibfnamefont {C.}~\bibnamefont
  {Genes}}, \ and\ \bibinfo {author} {\bibfnamefont {G.}~\bibnamefont
  {Pupillo}},\ }\href {https://arxiv.org/pdf/1801.09876.pdf
  http://arxiv.org/abs/1801.09876} {\  (\bibinfo {year} {2018})},\ \Eprint
  {http://arxiv.org/abs/1801.09876} {arXiv:1801.09876} \BibitemShut {NoStop}%
\bibitem [{\citenamefont {Martínez-Martínez}\ \emph
  {et~al.}(2018)\citenamefont {Martínez-Martínez}, \citenamefont {Du},
  \citenamefont {F.~Ribeiro}, \citenamefont {Kéna-Cohen},\ and\ \citenamefont
  {Yuen-Zhou}}]{Martinez-Martinez2017}%
  \BibitemOpen
  \bibfield  {author} {\bibinfo {author} {\bibfnamefont {L.~A.}\ \bibnamefont
  {Martínez-Martínez}}, \bibinfo {author} {\bibfnamefont {M.}~\bibnamefont
  {Du}}, \bibinfo {author} {\bibfnamefont {R.}~\bibnamefont {F.~Ribeiro}},
  \bibinfo {author} {\bibfnamefont {S.}~\bibnamefont {Kéna-Cohen}}, \ and\
  \bibinfo {author} {\bibfnamefont {J.}~\bibnamefont {Yuen-Zhou}},\ }\href
  {\doibase 10.1021/acs.jpclett.8b00008} {\bibfield  {journal} {\bibinfo
  {journal} {The Journal of Physical Chemistry Letters}\ }\textbf {\bibinfo
  {volume} {9}},\ \bibinfo {pages} {1951} (\bibinfo {year} {2018})}\BibitemShut
  {NoStop}%
\bibitem [{\citenamefont {Canaguier-Durand}\ \emph {et~al.}(2013)\citenamefont
  {Canaguier-Durand}, \citenamefont {Devaux}, \citenamefont {George},
  \citenamefont {Pang}, \citenamefont {Hutchison}, \citenamefont {Schwartz},
  \citenamefont {Genet}, \citenamefont {Wilhelms}, \citenamefont {Lehn},\ and\
  \citenamefont {Ebbesen}}]{Canaguier-durand2013}%
  \BibitemOpen
  \bibfield  {author} {\bibinfo {author} {\bibfnamefont {A.}~\bibnamefont
  {Canaguier-Durand}}, \bibinfo {author} {\bibfnamefont {E.}~\bibnamefont
  {Devaux}}, \bibinfo {author} {\bibfnamefont {J.}~\bibnamefont {George}},
  \bibinfo {author} {\bibfnamefont {Y.}~\bibnamefont {Pang}}, \bibinfo {author}
  {\bibfnamefont {J.~A.}\ \bibnamefont {Hutchison}}, \bibinfo {author}
  {\bibfnamefont {T.}~\bibnamefont {Schwartz}}, \bibinfo {author}
  {\bibfnamefont {C.}~\bibnamefont {Genet}}, \bibinfo {author} {\bibfnamefont
  {N.}~\bibnamefont {Wilhelms}}, \bibinfo {author} {\bibfnamefont {J.-M.}\
  \bibnamefont {Lehn}}, \ and\ \bibinfo {author} {\bibfnamefont {T.~W.}\
  \bibnamefont {Ebbesen}},\ }\href {\doibase 10.1002/anie.201301861} {\bibfield
   {journal} {\bibinfo  {journal} {Angew. Chemie Int. Ed.}\ }\textbf {\bibinfo
  {volume} {52}},\ \bibinfo {pages} {10533} (\bibinfo {year}
  {2013})}\BibitemShut {NoStop}%
\bibitem [{\citenamefont {Shalabney}\ \emph
  {et~al.}(2015{\natexlab{a}})\citenamefont {Shalabney}, \citenamefont
  {George}, \citenamefont {Hutchison}, \citenamefont {Pupillo}, \citenamefont
  {Genet},\ and\ \citenamefont {Ebbesen}}]{Shalabney15}%
  \BibitemOpen
  \bibfield  {author} {\bibinfo {author} {\bibfnamefont {A.}~\bibnamefont
  {Shalabney}}, \bibinfo {author} {\bibfnamefont {J.}~\bibnamefont {George}},
  \bibinfo {author} {\bibfnamefont {J.}~\bibnamefont {Hutchison}}, \bibinfo
  {author} {\bibfnamefont {G.}~\bibnamefont {Pupillo}}, \bibinfo {author}
  {\bibfnamefont {C.}~\bibnamefont {Genet}}, \ and\ \bibinfo {author}
  {\bibfnamefont {T.~W.}\ \bibnamefont {Ebbesen}},\ }\href {\doibase
  10.1038/ncomms6981} {\bibfield  {journal} {\bibinfo  {journal} {Nat. Comm.}\
  }\textbf {\bibinfo {volume} {6}},\ \bibinfo {pages} {5981} (\bibinfo {year}
  {2015}{\natexlab{a}})}\BibitemShut {NoStop}%
\bibitem [{\citenamefont {Galego}\ \emph {et~al.}(2015)\citenamefont {Galego},
  \citenamefont {Garcia-Vidal},\ and\ \citenamefont {Feist}}]{Galego15}%
  \BibitemOpen
  \bibfield  {author} {\bibinfo {author} {\bibfnamefont {J.}~\bibnamefont
  {Galego}}, \bibinfo {author} {\bibfnamefont {F.~J.}\ \bibnamefont
  {Garcia-Vidal}}, \ and\ \bibinfo {author} {\bibfnamefont {J.}~\bibnamefont
  {Feist}},\ }\href {\doibase 10.1103/PhysRevX.5.041022} {\bibfield  {journal}
  {\bibinfo  {journal} {Phys. Rev. X}\ }\textbf {\bibinfo {volume} {5}},\
  \bibinfo {pages} {41022} (\bibinfo {year} {2015})}\BibitemShut {NoStop}%
\bibitem [{\citenamefont {{\'{C}}wik}\ \emph {et~al.}(2016)\citenamefont
  {{\'{C}}wik}, \citenamefont {Kirton}, \citenamefont {{De Liberato}},\ and\
  \citenamefont {Keeling}}]{Cwik2016}%
  \BibitemOpen
  \bibfield  {author} {\bibinfo {author} {\bibfnamefont {J.~A.}\ \bibnamefont
  {{\'{C}}wik}}, \bibinfo {author} {\bibfnamefont {P.}~\bibnamefont {Kirton}},
  \bibinfo {author} {\bibfnamefont {S.}~\bibnamefont {{De Liberato}}}, \ and\
  \bibinfo {author} {\bibfnamefont {J.}~\bibnamefont {Keeling}},\ }\href
  {\doibase 10.1103/PhysRevA.93.033840} {\bibfield  {journal} {\bibinfo
  {journal} {Phys. Rev. A}\ }\textbf {\bibinfo {volume} {93}},\ \bibinfo
  {pages} {033840} (\bibinfo {year} {2016})}\BibitemShut {NoStop}%
\bibitem [{\citenamefont {Feist}\ \emph {et~al.}(2018)\citenamefont {Feist},
  \citenamefont {Galego},\ and\ \citenamefont {Garcia-Vidal}}]{Feist2018}%
  \BibitemOpen
  \bibfield  {author} {\bibinfo {author} {\bibfnamefont {J.}~\bibnamefont
  {Feist}}, \bibinfo {author} {\bibfnamefont {J.}~\bibnamefont {Galego}}, \
  and\ \bibinfo {author} {\bibfnamefont {F.~J.}\ \bibnamefont {Garcia-Vidal}},\
  }\href {\doibase 10.1021/acsphotonics.7b00680} {\bibfield  {journal}
  {\bibinfo  {journal} {ACS Photonics}\ }\textbf {\bibinfo {volume} {5}},\
  \bibinfo {pages} {205} (\bibinfo {year} {2018})}\BibitemShut {NoStop}%
\bibitem [{\citenamefont {Mart{\'{\i}}nez-Mart{\'{\i}}nez}\ \emph
  {et~al.}(2018)\citenamefont {Mart{\'{\i}}nez-Mart{\'{\i}}nez}, \citenamefont
  {Ribeiro}, \citenamefont {Campos-Gonz{\'{a}}lez-Angulo},\ and\ \citenamefont
  {Yuen-Zhou}}]{Martinez-Martinez2018}%
  \BibitemOpen
  \bibfield  {author} {\bibinfo {author} {\bibfnamefont {L.~A.}\ \bibnamefont
  {Mart{\'{\i}}nez-Mart{\'{\i}}nez}}, \bibinfo {author} {\bibfnamefont {R.~F.}\
  \bibnamefont {Ribeiro}}, \bibinfo {author} {\bibfnamefont {J.}~\bibnamefont
  {Campos-Gonz{\'{a}}lez-Angulo}}, \ and\ \bibinfo {author} {\bibfnamefont
  {J.}~\bibnamefont {Yuen-Zhou}},\ }\href {\doibase
  10.1021/acsphotonics.7b00610} {\bibfield  {journal} {\bibinfo  {journal} {ACS
  Photonics}\ }\textbf {\bibinfo {volume} {5}},\ \bibinfo {pages} {167}
  (\bibinfo {year} {2018})}\BibitemShut {NoStop}%
\bibitem [{\citenamefont {Bennett}\ \emph {et~al.}(2016)\citenamefont
  {Bennett}, \citenamefont {Kowalewski},\ and\ \citenamefont
  {Mukamel}}]{Bennett2016}%
  \BibitemOpen
  \bibfield  {author} {\bibinfo {author} {\bibfnamefont {K.}~\bibnamefont
  {Bennett}}, \bibinfo {author} {\bibfnamefont {M.}~\bibnamefont {Kowalewski}},
  \ and\ \bibinfo {author} {\bibfnamefont {S.}~\bibnamefont {Mukamel}},\ }\href
  {\doibase 10.1039/C6FD00095A} {\bibfield  {journal} {\bibinfo  {journal}
  {Faraday Discuss.}\ }\textbf {\bibinfo {volume} {194}},\ \bibinfo {pages}
  {259} (\bibinfo {year} {2016})}\BibitemShut {NoStop}%
\bibitem [{\citenamefont {Flick}\ \emph
  {et~al.}(2017{\natexlab{b}})\citenamefont {Flick}, \citenamefont {Appel},
  \citenamefont {Ruggenthaler},\ and\ \citenamefont {Rubio}}]{Flick2017b}%
  \BibitemOpen
  \bibfield  {author} {\bibinfo {author} {\bibfnamefont {J.}~\bibnamefont
  {Flick}}, \bibinfo {author} {\bibfnamefont {H.}~\bibnamefont {Appel}},
  \bibinfo {author} {\bibfnamefont {M.}~\bibnamefont {Ruggenthaler}}, \ and\
  \bibinfo {author} {\bibfnamefont {A.}~\bibnamefont {Rubio}},\ }\href
  {\doibase 10.1021/acs.jctc.6b01126} {\bibfield  {journal} {\bibinfo
  {journal} {J. Chem. Theory Comput.}\ }\textbf {\bibinfo {volume} {13}},\
  \bibinfo {pages} {1616} (\bibinfo {year} {2017}{\natexlab{b}})}\BibitemShut
  {NoStop}%
\bibitem [{\citenamefont {Herrera}\ and\ \citenamefont
  {Spano}(2018)}]{Herrera2018}%
  \BibitemOpen
  \bibfield  {author} {\bibinfo {author} {\bibfnamefont {F.}~\bibnamefont
  {Herrera}}\ and\ \bibinfo {author} {\bibfnamefont {F.~C.}\ \bibnamefont
  {Spano}},\ }\href {\doibase 10.1021/acsphotonics.7b00728} {\bibfield
  {journal} {\bibinfo  {journal} {ACS Photonics}\ }\textbf {\bibinfo {volume}
  {5}},\ \bibinfo {pages} {65} (\bibinfo {year} {2018})}\BibitemShut {NoStop}%
\bibitem [{\citenamefont {Kolaric}\ \emph {et~al.}(2018)\citenamefont
  {Kolaric}, \citenamefont {Maes}, \citenamefont {Clays}, \citenamefont
  {Durt},\ and\ \citenamefont {Caudano}}]{Kolaric2018}%
  \BibitemOpen
  \bibfield  {author} {\bibinfo {author} {\bibfnamefont {B.}~\bibnamefont
  {Kolaric}}, \bibinfo {author} {\bibfnamefont {B.}~\bibnamefont {Maes}},
  \bibinfo {author} {\bibfnamefont {K.}~\bibnamefont {Clays}}, \bibinfo
  {author} {\bibfnamefont {T.}~\bibnamefont {Durt}}, \ and\ \bibinfo {author}
  {\bibfnamefont {Y.}~\bibnamefont {Caudano}},\ }\href
  {http://arxiv.org/abs/1802.06029} {\  (\bibinfo {year} {2018})},\ \Eprint
  {http://arxiv.org/abs/1802.06029} {arXiv:1802.06029} \BibitemShut {NoStop}%
\bibitem [{\citenamefont {Ribeiro}\ \emph {et~al.}(2018)\citenamefont
  {Ribeiro}, \citenamefont {Mart{\'{\i}}nez-Mart{\'{\i}}nez}, \citenamefont
  {Du}, \citenamefont {Campos-Gonzalez-Angulo},\ and\ \citenamefont
  {Yuen-Zhou}}]{Ribeiro2018}%
  \BibitemOpen
  \bibfield  {author} {\bibinfo {author} {\bibfnamefont {R.~F.}\ \bibnamefont
  {Ribeiro}}, \bibinfo {author} {\bibfnamefont {L.~A.}\ \bibnamefont
  {Mart{\'{\i}}nez-Mart{\'{\i}}nez}}, \bibinfo {author} {\bibfnamefont
  {M.}~\bibnamefont {Du}}, \bibinfo {author} {\bibfnamefont {J.}~\bibnamefont
  {Campos-Gonzalez-Angulo}}, \ and\ \bibinfo {author} {\bibfnamefont
  {J.}~\bibnamefont {Yuen-Zhou}},\ }\href {http://arxiv.org/abs/1802.08681} {\
  (\bibinfo {year} {2018})},\ \Eprint {http://arxiv.org/abs/1802.08681}
  {arXiv:1802.08681} \BibitemShut {NoStop}%
\bibitem [{\citenamefont {Fausti}\ \emph {et~al.}(2011)\citenamefont {Fausti},
  \citenamefont {Tobey}, \citenamefont {Dean}, \citenamefont {Kaiser},
  \citenamefont {Dienst}, \citenamefont {Hoffmann}, \citenamefont {Pyon},
  \citenamefont {Takayama}, \citenamefont {Takagi},\ and\ \citenamefont
  {Cavalleri}}]{Fausti2011}%
  \BibitemOpen
  \bibfield  {author} {\bibinfo {author} {\bibfnamefont {D.}~\bibnamefont
  {Fausti}}, \bibinfo {author} {\bibfnamefont {R.~I.}\ \bibnamefont {Tobey}},
  \bibinfo {author} {\bibfnamefont {N.}~\bibnamefont {Dean}}, \bibinfo {author}
  {\bibfnamefont {S.}~\bibnamefont {Kaiser}}, \bibinfo {author} {\bibfnamefont
  {A.}~\bibnamefont {Dienst}}, \bibinfo {author} {\bibfnamefont {M.~C.}\
  \bibnamefont {Hoffmann}}, \bibinfo {author} {\bibfnamefont {S.}~\bibnamefont
  {Pyon}}, \bibinfo {author} {\bibfnamefont {T.}~\bibnamefont {Takayama}},
  \bibinfo {author} {\bibfnamefont {H.}~\bibnamefont {Takagi}}, \ and\ \bibinfo
  {author} {\bibfnamefont {A.}~\bibnamefont {Cavalleri}},\ }\href {\doibase
  10.1126/science.1197294} {\bibfield  {journal} {\bibinfo  {journal}
  {Science}\ }\textbf {\bibinfo {volume} {331}},\ \bibinfo {pages} {189}
  (\bibinfo {year} {2011})}\BibitemShut {NoStop}%
\bibitem [{\citenamefont {Mitrano}\ \emph {et~al.}(2016)\citenamefont
  {Mitrano}, \citenamefont {Cantaluppi}, \citenamefont {Nicoletti},
  \citenamefont {Kaiser}, \citenamefont {Perucchi}, \citenamefont {Lupi},
  \citenamefont {{Di Pietro}}, \citenamefont {Pontiroli}, \citenamefont
  {Ricc{\`{o}}}, \citenamefont {Clark}, \citenamefont {Jaksch},\ and\
  \citenamefont {Cavalleri}}]{Mitrano2016}%
  \BibitemOpen
  \bibfield  {author} {\bibinfo {author} {\bibfnamefont {M.}~\bibnamefont
  {Mitrano}}, \bibinfo {author} {\bibfnamefont {A.}~\bibnamefont {Cantaluppi}},
  \bibinfo {author} {\bibfnamefont {D.}~\bibnamefont {Nicoletti}}, \bibinfo
  {author} {\bibfnamefont {S.}~\bibnamefont {Kaiser}}, \bibinfo {author}
  {\bibfnamefont {A.}~\bibnamefont {Perucchi}}, \bibinfo {author}
  {\bibfnamefont {S.}~\bibnamefont {Lupi}}, \bibinfo {author} {\bibfnamefont
  {P.}~\bibnamefont {{Di Pietro}}}, \bibinfo {author} {\bibfnamefont
  {D.}~\bibnamefont {Pontiroli}}, \bibinfo {author} {\bibfnamefont
  {M.}~\bibnamefont {Ricc{\`{o}}}}, \bibinfo {author} {\bibfnamefont {S.~R.}\
  \bibnamefont {Clark}}, \bibinfo {author} {\bibfnamefont {D.}~\bibnamefont
  {Jaksch}}, \ and\ \bibinfo {author} {\bibfnamefont {A.}~\bibnamefont
  {Cavalleri}},\ }\href {\doibase 10.1038/nature16522} {\bibfield  {journal}
  {\bibinfo  {journal} {Nature}\ }\textbf {\bibinfo {volume} {530}},\ \bibinfo
  {pages} {461} (\bibinfo {year} {2016})}\BibitemShut {NoStop}%
\bibitem [{\citenamefont {Mankowsky}\ \emph {et~al.}(2014)\citenamefont
  {Mankowsky}, \citenamefont {Subedi}, \citenamefont {F{\"{o}}rst},
  \citenamefont {Mariager}, \citenamefont {Chollet}, \citenamefont {Lemke},
  \citenamefont {Robinson}, \citenamefont {Glownia}, \citenamefont {Minitti},
  \citenamefont {Frano}, \citenamefont {Fechner}, \citenamefont {Spaldin},
  \citenamefont {Loew}, \citenamefont {Keimer}, \citenamefont {Georges},\ and\
  \citenamefont {Cavalleri}}]{Mankowsky2014}%
  \BibitemOpen
  \bibfield  {author} {\bibinfo {author} {\bibfnamefont {R.}~\bibnamefont
  {Mankowsky}}, \bibinfo {author} {\bibfnamefont {A.}~\bibnamefont {Subedi}},
  \bibinfo {author} {\bibfnamefont {M.}~\bibnamefont {F{\"{o}}rst}}, \bibinfo
  {author} {\bibfnamefont {S.~O.}\ \bibnamefont {Mariager}}, \bibinfo {author}
  {\bibfnamefont {M.}~\bibnamefont {Chollet}}, \bibinfo {author} {\bibfnamefont
  {H.~T.}\ \bibnamefont {Lemke}}, \bibinfo {author} {\bibfnamefont {J.~S.}\
  \bibnamefont {Robinson}}, \bibinfo {author} {\bibfnamefont {J.~M.}\
  \bibnamefont {Glownia}}, \bibinfo {author} {\bibfnamefont {M.~P.}\
  \bibnamefont {Minitti}}, \bibinfo {author} {\bibfnamefont {A.}~\bibnamefont
  {Frano}}, \bibinfo {author} {\bibfnamefont {M.}~\bibnamefont {Fechner}},
  \bibinfo {author} {\bibfnamefont {N.~A.}\ \bibnamefont {Spaldin}}, \bibinfo
  {author} {\bibfnamefont {T.}~\bibnamefont {Loew}}, \bibinfo {author}
  {\bibfnamefont {B.}~\bibnamefont {Keimer}}, \bibinfo {author} {\bibfnamefont
  {A.}~\bibnamefont {Georges}}, \ and\ \bibinfo {author} {\bibfnamefont
  {A.}~\bibnamefont {Cavalleri}},\ }\href {\doibase 10.1038/nature13875}
  {\bibfield  {journal} {\bibinfo  {journal} {Nature}\ }\textbf {\bibinfo
  {volume} {516}},\ \bibinfo {pages} {71} (\bibinfo {year} {2014})}\BibitemShut
  {NoStop}%
\bibitem [{\citenamefont {Knap}\ \emph {et~al.}(2016)\citenamefont {Knap},
  \citenamefont {Babadi}, \citenamefont {Refael}, \citenamefont {Martin},\ and\
  \citenamefont {Demler}}]{Knap2016}%
  \BibitemOpen
  \bibfield  {author} {\bibinfo {author} {\bibfnamefont {M.}~\bibnamefont
  {Knap}}, \bibinfo {author} {\bibfnamefont {M.}~\bibnamefont {Babadi}},
  \bibinfo {author} {\bibfnamefont {G.}~\bibnamefont {Refael}}, \bibinfo
  {author} {\bibfnamefont {I.}~\bibnamefont {Martin}}, \ and\ \bibinfo {author}
  {\bibfnamefont {E.}~\bibnamefont {Demler}},\ }\href {\doibase
  10.1103/PhysRevB.94.214504} {\bibfield  {journal} {\bibinfo  {journal} {Phys.
  Rev. B}\ }\textbf {\bibinfo {volume} {94}},\ \bibinfo {pages} {214504}
  (\bibinfo {year} {2016})}\BibitemShut {NoStop}%
\bibitem [{\citenamefont {Kennes}\ \emph {et~al.}(2017)\citenamefont {Kennes},
  \citenamefont {Wilner}, \citenamefont {Reichman},\ and\ \citenamefont
  {Millis}}]{Kennes2017}%
  \BibitemOpen
  \bibfield  {author} {\bibinfo {author} {\bibfnamefont {D.~M.}\ \bibnamefont
  {Kennes}}, \bibinfo {author} {\bibfnamefont {E.~Y.}\ \bibnamefont {Wilner}},
  \bibinfo {author} {\bibfnamefont {D.~R.}\ \bibnamefont {Reichman}}, \ and\
  \bibinfo {author} {\bibfnamefont {A.~J.}\ \bibnamefont {Millis}},\ }\href
  {\doibase 10.1038/nphys4024} {\bibfield  {journal} {\bibinfo  {journal} {Nat.
  Phys.}\ }\textbf {\bibinfo {volume} {13}},\ \bibinfo {pages} {479} (\bibinfo
  {year} {2017})}\BibitemShut {NoStop}%
\bibitem [{\citenamefont {Sentef}\ \emph {et~al.}(2017)\citenamefont {Sentef},
  \citenamefont {Tokuno}, \citenamefont {Georges},\ and\ \citenamefont
  {Kollath}}]{sentef17}%
  \BibitemOpen
  \bibfield  {author} {\bibinfo {author} {\bibfnamefont {M.~A.}\ \bibnamefont
  {Sentef}}, \bibinfo {author} {\bibfnamefont {A.}~\bibnamefont {Tokuno}},
  \bibinfo {author} {\bibfnamefont {A.}~\bibnamefont {Georges}}, \ and\
  \bibinfo {author} {\bibfnamefont {C.}~\bibnamefont {Kollath}},\ }\href
  {\doibase 10.1103/PhysRevLett.118.087002} {\bibfield  {journal} {\bibinfo
  {journal} {Phys. Rev. Lett.}\ }\textbf {\bibinfo {volume} {118}},\ \bibinfo
  {pages} {087002} (\bibinfo {year} {2017})}\BibitemShut {NoStop}%
\bibitem [{\citenamefont {Schlawin}\ \emph {et~al.}(2017)\citenamefont
  {Schlawin}, \citenamefont {Dietrich}, \citenamefont {Kiffner}, \citenamefont
  {Cavalleri},\ and\ \citenamefont {Jaksch}}]{Schlawin2017}%
  \BibitemOpen
  \bibfield  {author} {\bibinfo {author} {\bibfnamefont {F.}~\bibnamefont
  {Schlawin}}, \bibinfo {author} {\bibfnamefont {A.~S.~D.}\ \bibnamefont
  {Dietrich}}, \bibinfo {author} {\bibfnamefont {M.}~\bibnamefont {Kiffner}},
  \bibinfo {author} {\bibfnamefont {A.}~\bibnamefont {Cavalleri}}, \ and\
  \bibinfo {author} {\bibfnamefont {D.}~\bibnamefont {Jaksch}},\ }\href
  {\doibase 10.1103/PhysRevB.96.064526} {\bibfield  {journal} {\bibinfo
  {journal} {Phys. Rev. B}\ }\textbf {\bibinfo {volume} {96}},\ \bibinfo
  {pages} {064526} (\bibinfo {year} {2017})}\BibitemShut {NoStop}%
\bibitem [{\citenamefont {Sentef}\ \emph {et~al.}(2018)\citenamefont {Sentef},
  \citenamefont {Ruggenthaler},\ and\ \citenamefont {Rubio}}]{Sentef2018}%
  \BibitemOpen
  \bibfield  {author} {\bibinfo {author} {\bibfnamefont {M.~A.}\ \bibnamefont
  {Sentef}}, \bibinfo {author} {\bibfnamefont {M.}~\bibnamefont
  {Ruggenthaler}}, \ and\ \bibinfo {author} {\bibfnamefont {A.}~\bibnamefont
  {Rubio}},\ }\href {http://arxiv.org/abs/1802.09437} {\  (\bibinfo {year}
  {2018})},\ \Eprint {http://arxiv.org/abs/1802.09437} {arXiv:1802.09437}
  \BibitemShut {NoStop}%
\bibitem [{\citenamefont {Schlawin}\ \emph {et~al.}(2018)\citenamefont
  {Schlawin}, \citenamefont {Cavalleri},\ and\ \citenamefont
  {Jaksch}}]{schlawin18}%
  \BibitemOpen
  \bibfield  {author} {\bibinfo {author} {\bibfnamefont {F.}~\bibnamefont
  {Schlawin}}, \bibinfo {author} {\bibfnamefont {A.}~\bibnamefont {Cavalleri}},
  \ and\ \bibinfo {author} {\bibfnamefont {D.}~\bibnamefont {Jaksch}},\
  }\href@noop {} {\  (\bibinfo {year} {2018})}\BibitemShut {NoStop}%
\bibitem [{\citenamefont {Wang}\ \emph {et~al.}(2014)\citenamefont {Wang},
  \citenamefont {Mika}, \citenamefont {Hutchison}, \citenamefont {Genet},
  \citenamefont {Jouaiti}, \citenamefont {Hosseini},\ and\ \citenamefont
  {Ebbesen}}]{Wang2014}%
  \BibitemOpen
  \bibfield  {author} {\bibinfo {author} {\bibfnamefont {S.}~\bibnamefont
  {Wang}}, \bibinfo {author} {\bibfnamefont {A.}~\bibnamefont {Mika}}, \bibinfo
  {author} {\bibfnamefont {J.~A.}\ \bibnamefont {Hutchison}}, \bibinfo {author}
  {\bibfnamefont {C.}~\bibnamefont {Genet}}, \bibinfo {author} {\bibfnamefont
  {A.}~\bibnamefont {Jouaiti}}, \bibinfo {author} {\bibfnamefont {M.~W.}\
  \bibnamefont {Hosseini}}, \ and\ \bibinfo {author} {\bibfnamefont {T.~W.}\
  \bibnamefont {Ebbesen}},\ }\href {\doibase 10.1039/C4NR01971G} {\bibfield
  {journal} {\bibinfo  {journal} {Nanoscale}\ }\textbf {\bibinfo {volume}
  {6}},\ \bibinfo {pages} {7243} (\bibinfo {year} {2014})}\BibitemShut
  {NoStop}%
\bibitem [{\citenamefont {George}\ \emph {et~al.}(2015)\citenamefont {George},
  \citenamefont {Shalabney}, \citenamefont {Hutchison}, \citenamefont {Genet},\
  and\ \citenamefont {Ebbesen}}]{George15}%
  \BibitemOpen
  \bibfield  {author} {\bibinfo {author} {\bibfnamefont {J.}~\bibnamefont
  {George}}, \bibinfo {author} {\bibfnamefont {A.}~\bibnamefont {Shalabney}},
  \bibinfo {author} {\bibfnamefont {J.~A.}\ \bibnamefont {Hutchison}}, \bibinfo
  {author} {\bibfnamefont {C.}~\bibnamefont {Genet}}, \ and\ \bibinfo {author}
  {\bibfnamefont {T.~W.}\ \bibnamefont {Ebbesen}},\ }\href {\doibase
  10.1021/acs.jpclett.5b00204} {\bibfield  {journal} {\bibinfo  {journal} {J.
  Phys. Chem. Lett.}\ }\textbf {\bibinfo {volume} {6}},\ \bibinfo {pages}
  {1027} (\bibinfo {year} {2015})}\BibitemShut {NoStop}%
\bibitem [{\citenamefont {del Pino}\ \emph
  {et~al.}(2015{\natexlab{a}})\citenamefont {del Pino}, \citenamefont {Feist},\
  and\ \citenamefont {Garcia-Vidal}}]{Pino15}%
  \BibitemOpen
  \bibfield  {author} {\bibinfo {author} {\bibfnamefont {J.}~\bibnamefont {del
  Pino}}, \bibinfo {author} {\bibfnamefont {J.}~\bibnamefont {Feist}}, \ and\
  \bibinfo {author} {\bibfnamefont {F.~J.}\ \bibnamefont {Garcia-Vidal}},\
  }\href {\doibase 10.1088/1367-2630/17/5/053040} {\bibfield  {journal}
  {\bibinfo  {journal} {New J. Phys.}\ }\textbf {\bibinfo {volume} {17}},\
  \bibinfo {pages} {53040} (\bibinfo {year} {2015}{\natexlab{a}})}\BibitemShut
  {NoStop}%
\bibitem [{\citenamefont {Shalabney}\ \emph
  {et~al.}(2015{\natexlab{b}})\citenamefont {Shalabney}, \citenamefont
  {George}, \citenamefont {Hiura}, \citenamefont {Hutchison}, \citenamefont
  {Genet}, \citenamefont {Hellwig}, \citenamefont {Ebbesen}, \citenamefont
  {Hellwig},\ and\ \citenamefont {Ebbesen}}]{shalabney15:raman}%
  \BibitemOpen
  \bibfield  {author} {\bibinfo {author} {\bibfnamefont {A.}~\bibnamefont
  {Shalabney}}, \bibinfo {author} {\bibfnamefont {J.}~\bibnamefont {George}},
  \bibinfo {author} {\bibfnamefont {H.}~\bibnamefont {Hiura}}, \bibinfo
  {author} {\bibfnamefont {J.~a.}\ \bibnamefont {Hutchison}}, \bibinfo {author}
  {\bibfnamefont {C.}~\bibnamefont {Genet}}, \bibinfo {author} {\bibfnamefont
  {P.}~\bibnamefont {Hellwig}}, \bibinfo {author} {\bibfnamefont {T.~W.}\
  \bibnamefont {Ebbesen}}, \bibinfo {author} {\bibfnamefont {P.}~\bibnamefont
  {Hellwig}}, \ and\ \bibinfo {author} {\bibfnamefont {T.~W.}\ \bibnamefont
  {Ebbesen}},\ }\href {\doibase 10.1002/anie.201502979} {\bibfield  {journal}
  {\bibinfo  {journal} {Ang. Chem. Int. Ed.}\ }\textbf {\bibinfo {volume}
  {54}},\ \bibinfo {pages} {7971} (\bibinfo {year}
  {2015}{\natexlab{b}})}\BibitemShut {NoStop}%
\bibitem [{\citenamefont {del Pino}\ \emph
  {et~al.}(2015{\natexlab{b}})\citenamefont {del Pino}, \citenamefont {Feist},\
  and\ \citenamefont {Garcia-Vidal}}]{Pino15a}%
  \BibitemOpen
  \bibfield  {author} {\bibinfo {author} {\bibfnamefont {J.}~\bibnamefont {del
  Pino}}, \bibinfo {author} {\bibfnamefont {J.}~\bibnamefont {Feist}}, \ and\
  \bibinfo {author} {\bibfnamefont {F.~J.}\ \bibnamefont {Garcia-Vidal}},\
  }\href {\doibase 10.1021/acs.jpcc.5b11654} {\bibfield  {journal} {\bibinfo
  {journal} {J. Phys. Chem. C}\ }\textbf {\bibinfo {volume} {119}},\ \bibinfo
  {pages} {29132} (\bibinfo {year} {2015}{\natexlab{b}})}\BibitemShut {NoStop}%
\bibitem [{\citenamefont {Strashko}\ and\ \citenamefont
  {Keeling}(2016)}]{strashko2016raman}%
  \BibitemOpen
  \bibfield  {author} {\bibinfo {author} {\bibfnamefont {A.}~\bibnamefont
  {Strashko}}\ and\ \bibinfo {author} {\bibfnamefont {J.}~\bibnamefont
  {Keeling}},\ }\href {\doibase 10.1103/PhysRevA.94.023843} {\bibfield
  {journal} {\bibinfo  {journal} {Phys. Rev. A}\ }\textbf {\bibinfo {volume}
  {94}},\ \bibinfo {pages} {023843} (\bibinfo {year} {2016})}\BibitemShut
  {NoStop}%
\bibitem [{\citenamefont {Flick}\ \emph {et~al.}(2018)\citenamefont {Flick},
  \citenamefont {Schäfer}, \citenamefont {Ruggenthaler}, \citenamefont
  {Appel},\ and\ \citenamefont {Rubio}}]{Flick2017}%
  \BibitemOpen
  \bibfield  {author} {\bibinfo {author} {\bibfnamefont {J.}~\bibnamefont
  {Flick}}, \bibinfo {author} {\bibfnamefont {C.}~\bibnamefont {Schäfer}},
  \bibinfo {author} {\bibfnamefont {M.}~\bibnamefont {Ruggenthaler}}, \bibinfo
  {author} {\bibfnamefont {H.}~\bibnamefont {Appel}}, \ and\ \bibinfo {author}
  {\bibfnamefont {A.}~\bibnamefont {Rubio}},\ }\href {\doibase
  10.1021/acsphotonics.7b01279} {\bibfield  {journal} {\bibinfo  {journal} {ACS
  photonics}\ }\textbf {\bibinfo {volume} {5}},\ \bibinfo {pages} {992}
  (\bibinfo {year} {2018})}\BibitemShut {NoStop}%
\bibitem [{\citenamefont {Herrera}\ and\ \citenamefont
  {Spano}(2017{\natexlab{a}})}]{Herrera2016a}%
  \BibitemOpen
  \bibfield  {author} {\bibinfo {author} {\bibfnamefont {F.}~\bibnamefont
  {Herrera}}\ and\ \bibinfo {author} {\bibfnamefont {F.~C.}\ \bibnamefont
  {Spano}},\ }\href {\doibase 10.1103/PhysRevLett.118.223601} {\bibfield
  {journal} {\bibinfo  {journal} {Phys. Rev. Lett.}\ }\textbf {\bibinfo
  {volume} {118}},\ \bibinfo {pages} {223601} (\bibinfo {year}
  {2017}{\natexlab{a}})}\BibitemShut {NoStop}%
\bibitem [{\citenamefont {Herrera}\ and\ \citenamefont
  {Spano}(2017{\natexlab{b}})}]{Herrera2016}%
  \BibitemOpen
  \bibfield  {author} {\bibinfo {author} {\bibfnamefont {F.}~\bibnamefont
  {Herrera}}\ and\ \bibinfo {author} {\bibfnamefont {F.~C.}\ \bibnamefont
  {Spano}},\ }\href {\doibase 10.1103/PhysRevA.95.053867} {\bibfield  {journal}
  {\bibinfo  {journal} {Phys. Rev. A}\ }\textbf {\bibinfo {volume} {95}},\
  \bibinfo {pages} {053867} (\bibinfo {year} {2017}{\natexlab{b}})}\BibitemShut
  {NoStop}%
\bibitem [{\citenamefont {Cortese}\ \emph {et~al.}(2017)\citenamefont
  {Cortese}, \citenamefont {Lagoudakis},\ and\ \citenamefont {{De
  Liberato}}}]{Cortese2017}%
  \BibitemOpen
  \bibfield  {author} {\bibinfo {author} {\bibfnamefont {E.}~\bibnamefont
  {Cortese}}, \bibinfo {author} {\bibfnamefont {P.~G.}\ \bibnamefont
  {Lagoudakis}}, \ and\ \bibinfo {author} {\bibfnamefont {S.}~\bibnamefont {{De
  Liberato}}},\ }\href {\doibase 10.1103/PhysRevLett.119.043604} {\bibfield
  {journal} {\bibinfo  {journal} {Phys. Rev. Lett.}\ }\textbf {\bibinfo
  {volume} {119}},\ \bibinfo {pages} {043604} (\bibinfo {year}
  {2017})}\BibitemShut {NoStop}%
\bibitem [{\citenamefont {Zeb}\ \emph {et~al.}(2018)\citenamefont {Zeb},
  \citenamefont {Kirton},\ and\ \citenamefont {Keeling}}]{Zeb2017}%
  \BibitemOpen
  \bibfield  {author} {\bibinfo {author} {\bibfnamefont {M.~A.}\ \bibnamefont
  {Zeb}}, \bibinfo {author} {\bibfnamefont {P.~G.}\ \bibnamefont {Kirton}}, \
  and\ \bibinfo {author} {\bibfnamefont {J.}~\bibnamefont {Keeling}},\ }\href
  {\doibase 10.1021/acsphotonics.7b00916} {\bibfield  {journal} {\bibinfo
  {journal} {ACS Photonics}\ }\textbf {\bibinfo {volume} {5}},\ \bibinfo
  {pages} {249} (\bibinfo {year} {2018})}\BibitemShut {NoStop}%
\bibitem [{\citenamefont {Wang}\ and\ \citenamefont {Hioe}(1973)}]{wang73}%
  \BibitemOpen
  \bibfield  {author} {\bibinfo {author} {\bibfnamefont {Y.~K.}\ \bibnamefont
  {Wang}}\ and\ \bibinfo {author} {\bibfnamefont {F.~T.}\ \bibnamefont
  {Hioe}},\ }\href {\doibase 10.1103/PhysRevA.7.831} {\bibfield  {journal}
  {\bibinfo  {journal} {Phys. Rev. A}\ }\textbf {\bibinfo {volume} {7}},\
  \bibinfo {pages} {831} (\bibinfo {year} {1973})}\BibitemShut {NoStop}%
\bibitem [{\citenamefont {Eastham}\ and\ \citenamefont
  {Littlewood}(2001)}]{eastham01}%
  \BibitemOpen
  \bibfield  {author} {\bibinfo {author} {\bibfnamefont {P.~R.}\ \bibnamefont
  {Eastham}}\ and\ \bibinfo {author} {\bibfnamefont {P.~B.}\ \bibnamefont
  {Littlewood}},\ }\href {\doibase 10.1103/PhysRevB.64.235101} {\bibfield
  {journal} {\bibinfo  {journal} {Phys. Rev. B}\ }\textbf {\bibinfo {volume}
  {64}},\ \bibinfo {pages} {235101} (\bibinfo {year} {2001})}\BibitemShut
  {NoStop}%
\bibitem [{\citenamefont {Touchette}(2009)}]{Touchette2009}%
  \BibitemOpen
  \bibfield  {author} {\bibinfo {author} {\bibfnamefont {H.}~\bibnamefont
  {Touchette}},\ }\href {\doibase 10.1016/j.physrep.2009.05.002} {\bibfield
  {journal} {\bibinfo  {journal} {Phys. Rep.}\ }\textbf {\bibinfo {volume}
  {478}},\ \bibinfo {pages} {1} (\bibinfo {year} {2009})}\BibitemShut {NoStop}%
\bibitem [{\citenamefont {Garraway}(2011)}]{Garraway2011a}%
  \BibitemOpen
  \bibfield  {author} {\bibinfo {author} {\bibfnamefont {B.~M.}\ \bibnamefont
  {Garraway}},\ }\href {\doibase 10.1098/rsta.2010.0333} {\bibfield  {journal}
  {\bibinfo  {journal} {Phil. Trans. R. Soc. A}\ }\textbf {\bibinfo {volume}
  {369}},\ \bibinfo {pages} {1137} (\bibinfo {year} {2011})}\BibitemShut
  {NoStop}%
\bibitem [{\citenamefont {Mivehvar}\ \emph {et~al.}(2017)\citenamefont
  {Mivehvar}, \citenamefont {Piazza},\ and\ \citenamefont
  {Ritsch}}]{Mivehvar2017}%
  \BibitemOpen
  \bibfield  {author} {\bibinfo {author} {\bibfnamefont {F.}~\bibnamefont
  {Mivehvar}}, \bibinfo {author} {\bibfnamefont {F.}~\bibnamefont {Piazza}}, \
  and\ \bibinfo {author} {\bibfnamefont {H.}~\bibnamefont {Ritsch}},\ }\href
  {\doibase 10.1103/PhysRevLett.119.063602} {\bibfield  {journal} {\bibinfo
  {journal} {Phys. Rev. Lett.}\ }\textbf {\bibinfo {volume} {119}},\ \bibinfo
  {pages} {063602} (\bibinfo {year} {2017})}\BibitemShut {NoStop}%
\bibitem [{\citenamefont {Kakuyanagi}\ \emph {et~al.}(2016)\citenamefont
  {Kakuyanagi}, \citenamefont {Matsuzaki}, \citenamefont {D{\'{e}}prez},
  \citenamefont {Toida}, \citenamefont {Semba}, \citenamefont {Yamaguchi},
  \citenamefont {Munro},\ and\ \citenamefont {Saito}}]{Kakuyanagi2016}%
  \BibitemOpen
  \bibfield  {author} {\bibinfo {author} {\bibfnamefont {K.}~\bibnamefont
  {Kakuyanagi}}, \bibinfo {author} {\bibfnamefont {Y.}~\bibnamefont
  {Matsuzaki}}, \bibinfo {author} {\bibfnamefont {C.}~\bibnamefont
  {D{\'{e}}prez}}, \bibinfo {author} {\bibfnamefont {H.}~\bibnamefont {Toida}},
  \bibinfo {author} {\bibfnamefont {K.}~\bibnamefont {Semba}}, \bibinfo
  {author} {\bibfnamefont {H.}~\bibnamefont {Yamaguchi}}, \bibinfo {author}
  {\bibfnamefont {W.~J.}\ \bibnamefont {Munro}}, \ and\ \bibinfo {author}
  {\bibfnamefont {S.}~\bibnamefont {Saito}},\ }\href {\doibase
  10.1103/PhysRevLett.117.210503} {\bibfield  {journal} {\bibinfo  {journal}
  {Phys. Rev. Lett.}\ }\textbf {\bibinfo {volume} {117}},\ \bibinfo {pages}
  {210503} (\bibinfo {year} {2016})}\BibitemShut {NoStop}%
\bibitem [{\citenamefont {Marchetti}\ \emph {et~al.}(2007)\citenamefont
  {Marchetti}, \citenamefont {Keeling}, \citenamefont {Szyma\'{n}ska},\ and\
  \citenamefont {Littlewood}}]{Marchetti2007a}%
  \BibitemOpen
  \bibfield  {author} {\bibinfo {author} {\bibfnamefont {F.~M.}\ \bibnamefont
  {Marchetti}}, \bibinfo {author} {\bibfnamefont {J.}~\bibnamefont {Keeling}},
  \bibinfo {author} {\bibfnamefont {M.~H.}\ \bibnamefont {Szyma\'{n}ska}}, \
  and\ \bibinfo {author} {\bibfnamefont {P.~B.}\ \bibnamefont {Littlewood}},\
  }\href {\doibase 10.1103/PhysRevB.76.115326} {\bibfield  {journal} {\bibinfo
  {journal} {Phys. Rev. B}\ }\textbf {\bibinfo {volume} {76}},\ \bibinfo
  {pages} {115326} (\bibinfo {year} {2007})}\BibitemShut {NoStop}%
\bibitem [{\citenamefont {{\'{C}}wik}\ \emph {et~al.}(2014)\citenamefont
  {{\'{C}}wik}, \citenamefont {Reja}, \citenamefont {Littlewood},\ and\
  \citenamefont {Keeling}}]{Cwik2014}%
  \BibitemOpen
  \bibfield  {author} {\bibinfo {author} {\bibfnamefont {J.~A.}\ \bibnamefont
  {{\'{C}}wik}}, \bibinfo {author} {\bibfnamefont {S.}~\bibnamefont {Reja}},
  \bibinfo {author} {\bibfnamefont {P.~B.}\ \bibnamefont {Littlewood}}, \ and\
  \bibinfo {author} {\bibfnamefont {J.}~\bibnamefont {Keeling}},\ }\href
  {\doibase 10.1209/0295-5075/105/47009} {\bibfield  {journal} {\bibinfo
  {journal} {Eur. Lett.}\ }\textbf {\bibinfo {volume} {105}},\ \bibinfo {pages}
  {47009} (\bibinfo {year} {2014})}\BibitemShut {NoStop}%
\bibitem [{\citenamefont {Moodie}\ \emph {et~al.}(2017)\citenamefont {Moodie},
  \citenamefont {Kirton},\ and\ \citenamefont {Keeling}}]{Moodie2017}%
  \BibitemOpen
  \bibfield  {author} {\bibinfo {author} {\bibfnamefont {R.~I.}\ \bibnamefont
  {Moodie}}, \bibinfo {author} {\bibfnamefont {P.}~\bibnamefont {Kirton}}, \
  and\ \bibinfo {author} {\bibfnamefont {J.}~\bibnamefont {Keeling}},\ }\href
  {\doibase 10.1103/PhysRevA.96.043844} {\bibfield  {journal} {\bibinfo
  {journal} {Phys. Rev. A}\ }\textbf {\bibinfo {volume} {96}},\ \bibinfo
  {pages} {043844} (\bibinfo {year} {2017})}\BibitemShut {NoStop}%
\bibitem [{\citenamefont {Greveling}\ \emph {et~al.}(2017)\citenamefont
  {Greveling}, \citenamefont {van~der Laan}, \citenamefont {Jagers},\ and\
  \citenamefont {van Oosten}}]{Greveling2017}%
  \BibitemOpen
  \bibfield  {author} {\bibinfo {author} {\bibfnamefont {S.}~\bibnamefont
  {Greveling}}, \bibinfo {author} {\bibfnamefont {F.}~\bibnamefont {van~der
  Laan}}, \bibinfo {author} {\bibfnamefont {H.~C.}\ \bibnamefont {Jagers}}, \
  and\ \bibinfo {author} {\bibfnamefont {D.}~\bibnamefont {van Oosten}},\
  }\href {http://arxiv.org/abs/1712.08426} {\  (\bibinfo {year} {2017})},\
  \Eprint {http://arxiv.org/abs/1712.08426} {arXiv:1712.08426} \BibitemShut
  {NoStop}%
\end{thebibliography}
%

\end{document}